\documentclass[journal,web]{ieeecolor}
\usepackage{generic} 
\usepackage{subfiles}
\usepackage{cite}
\usepackage{makeidx}
\usepackage{latexsym}
\usepackage{amsmath,amssymb,amsfonts}
\usepackage{threeparttable}
\usepackage{adjustbox}
\usepackage{graphicx}
\usepackage{textcomp}
\usepackage{multirow}
\usepackage[svgnames]{xcolor}
\usepackage[colorlinks]{hyperref}
\hypersetup{linkcolor=DarkRed}
\usepackage{cleveref}
\usepackage{verbatim}
\usepackage{epsfig}
\usepackage{array}
\usepackage{marginnote}

\definecolor{red}{rgb}{0.8,0,0}
\definecolor{blue}{rgb}{0,0,0.8}
\definecolor{green}{rgb}{0,0.4,0}

\newcommand{\change}[2]{}
\newcommand{\lchange}[2]{}

\def\BibTeX{{\rm B\kern-.05em{\sc i\kern-.025em b}\kern-.08em
    T\kern-.1667em\lower.7ex\hbox{E}\kern-.125emX}}
\begin{document}

\clearpage

\twocolumn
\pagenumbering{arabic}
\setcounter{page}{1}

\title{Learning Tubule-Sensitive CNNs for Pulmonary Airway and Artery-Vein Segmentation in CT}
\author{Yulei Qin, \IEEEmembership{Student Member, IEEE}, Hao Zheng, Yun Gu, \IEEEmembership{Member, IEEE}, Xiaolin Huang, \IEEEmembership{Senior Member, IEEE}, Jie Yang, Lihui Wang, Feng Yao, Yue-Min Zhu, and Guang-Zhong Yang, \IEEEmembership{Fellow, IEEE}
\thanks{This work was partly supported by National Natural Science Foundation of China (Nos. 62003208, 61661010, 61977046), National Key R\&D Program of China (No. 2019YFB1311503), Committee of Science and Technology, Shanghai, China (Nos. 19411963900, 19510711200), Shanghai Sailing Program (No. 20YF1420800), Shanghai Municipal of Science and Technology (Project No. 20JC1419500, Major Project No. 2021SHZDZX0102), International Research Project METISLAB, Program PHC-Cai Yuanpei 2018 (No. 41400TC), and China Scholarship Council (No. 201906230173).
\emph{(Corresponding authors: Jie Yang, Yun Gu, and Guang-Zhong Yang.)}}
\thanks{Y. Qin, H. Zheng, Y. Gu, X. Huang, and J. Yang are with the Institute of Image Processing and Pattern Recognition, Institute of Medical Robotics, Shanghai Jiao Tong University, Shanghai, China (e-mail: \{jieyang, geron762\}@sjtu.edu.cn).}
\thanks{L. Wang is with the Key Laboratory of Intelligent Medical Image Analysis and Precise Diagnosis of Guizhou Province, School of Computer Science and Technology, Guizhou University, Guiyang, China.}
\thanks{F. Yao is with the Department of Thoracic Surgery, Shanghai Chest Hospital, Shanghai Jiao Tong University, Shanghai, China.}
\thanks{Y. Qin and Y.-M. Zhu are with the Universit\'{e} de Lyon, INSA Lyon, CREATIS, CNRS, INSERM, UMR 5220, U1206, Villeurbanne, France.}
\thanks{G.-Z. Yang is with the Institute of Medical Robotics, School of Biomedical Engineering, Shanghai Jiao Tong University, Shanghai, China (e-mail: gzyang@sjtu.edu.cn).}
}

\maketitle

\begin{abstract}
Training convolutional neural networks (CNNs) for segmentation of pulmonary airway, artery, and vein is challenging due to sparse supervisory signals caused by the severe class imbalance between tubular targets and background. We present a CNNs-based method for accurate airway and artery-vein segmentation in non-contrast computed tomography. It enjoys superior sensitivity to tenuous peripheral bronchioles, arterioles, and venules. The method first uses a feature recalibration module to make the best use of features learned from the neural networks. Spatial information of features is properly integrated to retain relative priority of activated regions, which benefits the subsequent channel-wise recalibration. Then, attention distillation module is introduced to reinforce representation learning of tubular objects. Fine-grained details in high-resolution attention maps are passing down from one layer to its previous layer recursively to enrich context. Anatomy prior of lung context map and distance transform map is designed and incorporated for better artery-vein differentiation capacity. Extensive experiments demonstrated considerable performance gains brought by these components. Compared with state-of-the-art methods, our method extracted much more branches while maintaining competitive overall segmentation performance. Codes and models are available at \url{http://www.pami.sjtu.edu.cn/News/56}.
\end{abstract}


\section{Introduction}
\label{sec:introduction}
\IEEEPARstart{P}{ulmonary} diseases pose high risks to human health. As a diagnostic tool, computed tomography (CT) has been widely adopted to reveal tomographic patterns of pulmonary diseases. It is of significant clinical interest to study pulmonary structures in volume-of-interest (VOI). One prerequisite step is to extract pulmonary airways from CT. The modeling of airway tree benefits the quantification of its morphological changes for diagnosis of bronchial stenosis, acute respiratory distress syndrome, idiopathic pulmonary fibrosis, chronic obstructive pulmonary diseases (COPD), obliterative bronchiolitis, and pulmonary contusion \cite{howling1998significance, shaw2002role, fetita2004pulmonary, li2019application, wu2019computed}. Combined with photo-realistic rendering and projection, the segmented airways play an important role in virtual bronchoscopy and endobronchial navigation for surgery \cite{mori2000automated, natori2005virtual, shen2015robust, shen2019context}. Another essential step is to extract pulmonary arteries and veins from CT. Pulmonary diseases may affect artery or vein, or both but in different ways \cite{melot2011pulmonary, charbonnier2015automatic}. Morphological changes of arteries are measured in diagnosing pulmonary embolism, arteriovenous malformations, and COPD \cite{zhou2007automatic, wittenberg2012acute, cartin2013pulmonary, estepar2013computed}. The arterial alterations also serve as an imaging biomarker in chronic thromboembolic pulmonary hypertension \cite{rahaghi2016pulmonary}. Accurate separation of veins from arteries may improve computer-aided diagnosis of embolism because most false positive lesions were found in veins \cite{wittenberg2012acute}. The imaging features of veins are found useful in diagnosis of vein diseases \cite{porres2013learning}. Despite the benefits of airway and artery-vein segmentation, it requires heavy workloads for manual delineation due to the complexity of tubular structures. Consequently, automatic segmentation methods were developed to reduce burden and improve accuracy. Especially if arteries and veins can be extracted from non-contrast CT (i.e. CT without the use of contrast agents), CT pulmonary angiogram may not be needed in certain cases to avoid adverse reactions to contrast agents \cite{cochran2001trends, loh2010delayed}.


\subsection{Related Work}


Over the past decades, several methods have been proposed for airway segmentation \cite{mori2000automated, van2009automatic, lo2012extraction}. Most of them employed techniques such as adaptive thresholding, region growing and filtering-based enhancement. These methods successfully segmented thick bronchi, but often failed to extract peripheral bronchioles due to the fact that the intensity contrast between airway lumen and wall weakens as airways bifurcate into thinner branches. Recent progress of convolutional neural networks (CNNs) has spawned research on airway segmentation using CNNs \cite{selvan2020graph, juarez2019joint, wang2019tubular, qin2019airwaynet, yun2019improvement, charbonnier2017improving, meng2017tracking, jin20173d, juarez2018automatic, zhao2019bronchus}. Two-dimensional (2-D) and 2.5-D CNNs \cite{yun2019improvement, charbonnier2017improving} were respectively applied on the initial coarsely segmented bronchi to reduce false positives and increase length of the detected airway tree. Three-dimensional (3-D) CNNs were developed for direct airway segmentation in either a dynamic VOI-based tracking way \cite{meng2017tracking} or a fixed-stride sliding window way \cite{juarez2018automatic}. The spatial recurrent convolution layer and radial distance loss were proposed by \cite{wang2019tubular} for tubular topology perception. In \cite{qin2019airwaynet}, the airway segmentation task was transformed into 26-neighbor connectivity prediction task for inherent structure comprehension. Both 2-D and 3-D CNNs were combined with linear programming-based tracking in \cite{zhao2019bronchus}. Graph neural networks \cite{selvan2020graph, juarez2019joint} were explored to incorporate neighborhood knowledge in feature utilization.


Previous methods on artery-vein separation relied on the enhanced or segmented vessels as premise \cite{buelow2005automatic, mekada2006pulmonary, saha2010topomorphologic, gao2012new, payer2016automated, charbonnier2015automatic, nardelli2018pulmonary}. To tackle the variety of vessels, combination of techniques such as local filtering and anatomical guidance is employed in the literature. Specifically, they utilized the proximity of airways to arteries for differentiation and suppressed airway walls to reduce false positives. Buelow \emph{et al.} \cite{buelow2005automatic} proposed a measure of ``arterialness" by identifying airway candidates in the vicinity of given vessels and assigning high value to vessels that run in parallel with bronchi. Mekada \emph{et al.} \cite{mekada2006pulmonary} calculated the distance from vessels to airways and to inter-lobar fissures. Vessels closer to airways are arteries and those closer to fissures are veins. Both Saha \emph{et al.} \cite{saha2010topomorphologic} and Gao \emph{et al.} \cite{gao2012new} combined distance transform and fuzzy connectivity with morphological opening for separation. Recently, three methods were developed to improve artery-vein segmentation \cite{charbonnier2015automatic, payer2016automated, nardelli2018pulmonary}. Charbonnier \emph{et al.} \cite{charbonnier2015automatic} first constructed a graph representation of the segmented vessels to extract sub-trees. These trees were grouped iteratively and the final classification was performed by comparing the volume size of the linked trees. Payer \emph{et al.} \cite{payer2016automated} extracted vessel sub-trees and classified each sub-tree via integer programming. Two anatomy properties were used: 1) proximity of arteries to bronchi; 2) uniform distribution of arteries and veins. CNNs were at the first time introduced to artery-vein classification by Nardelli \emph{et al.} \cite{nardelli2018pulmonary}. Graph-cut was adopted as post-processing to remove spatial inconsistency.

\subsection{Limitations and Challenges}
\label{sec:limitandchallenge}
Despite the improved performance of pulmonary airway and artery-vein segmentation by deep learning, there still remain limitations and challenges to be overcome.

First, for both airway and artery-vein segmentation, the severe class imbalance between tubular foreground and background poses a threat to the training of 3-D CNNs. Most CNNs heavily rely on airway and vessel ground-truth as supervisory signals. Unlike bulky or spheroid-like organs (e.g., liver and kidney), tree-like airways, arteries, and veins are thin, tenuous and divergent. The number of annotated voxels are far fewer than that of background voxels in the thoracic cavity. It is difficult to train deep models using such sparse and scattered targets. Although weighted cross-entropy loss and data sampling strategies were proposed to focus on the minority, single source of supervisory signals from deficient airway and artery-vein labels still makes optimization ineffective.

Second, the spatial distribution and branching pattern of airways and vessels require the model to utilize both global-scale and local-scale context to perceive the main body (e.g., trachea, main branches) and limbs (e.g., peripheral bronchi and vessels). Previous deep learning models used 2 or 3 pooling layers and the coarsest resolution features provide limited long-range context. If more layers are simply piled up, the increased parameters may cause over-fitting due to inadequate training data. If the width of CNNs (a.k.a. number of feature channels) is sacrificed for the depth (a.k.a. number of convolution layers) to avoid such parameter ``explosion", the model's learning and fitting capacity may get restricted.

Third, it is more rigorous to deem pulmonary artery-vein separation methods in the literature as classification rather than segmentation. They used two-stage strategy and counted on vessel segmentation in the first stage. The subsequent artery-vein separation was treated as an independent classification task in the second stage. Different techniques were deployed in two stages and such isolation has two drawbacks: 1) It blocks the path for the second model to exploit rich context from the first one, especially when CNNs are applied as backbone. CNNs cannot take advantage of high correlation between the two tasks and have to learn from scratch. 2) The performance of artery-vein segmentation is largely affected by that of vessel extraction and errors accumulate along the whole pipeline.

Last but not least, for artery-vein separation, previous CNNs-based method \cite{nardelli2018pulmonary} did not consider the relationship between airways, arteries, and veins. Auxiliary anatomy prior (e.g., close proximity of arteries to airways, intensity similarity between airway walls and vessels) was not involved in algorithm design, leaving room for further improvement.

\begin{figure*}[htbp]
\centerline{\includegraphics[width=\textwidth]{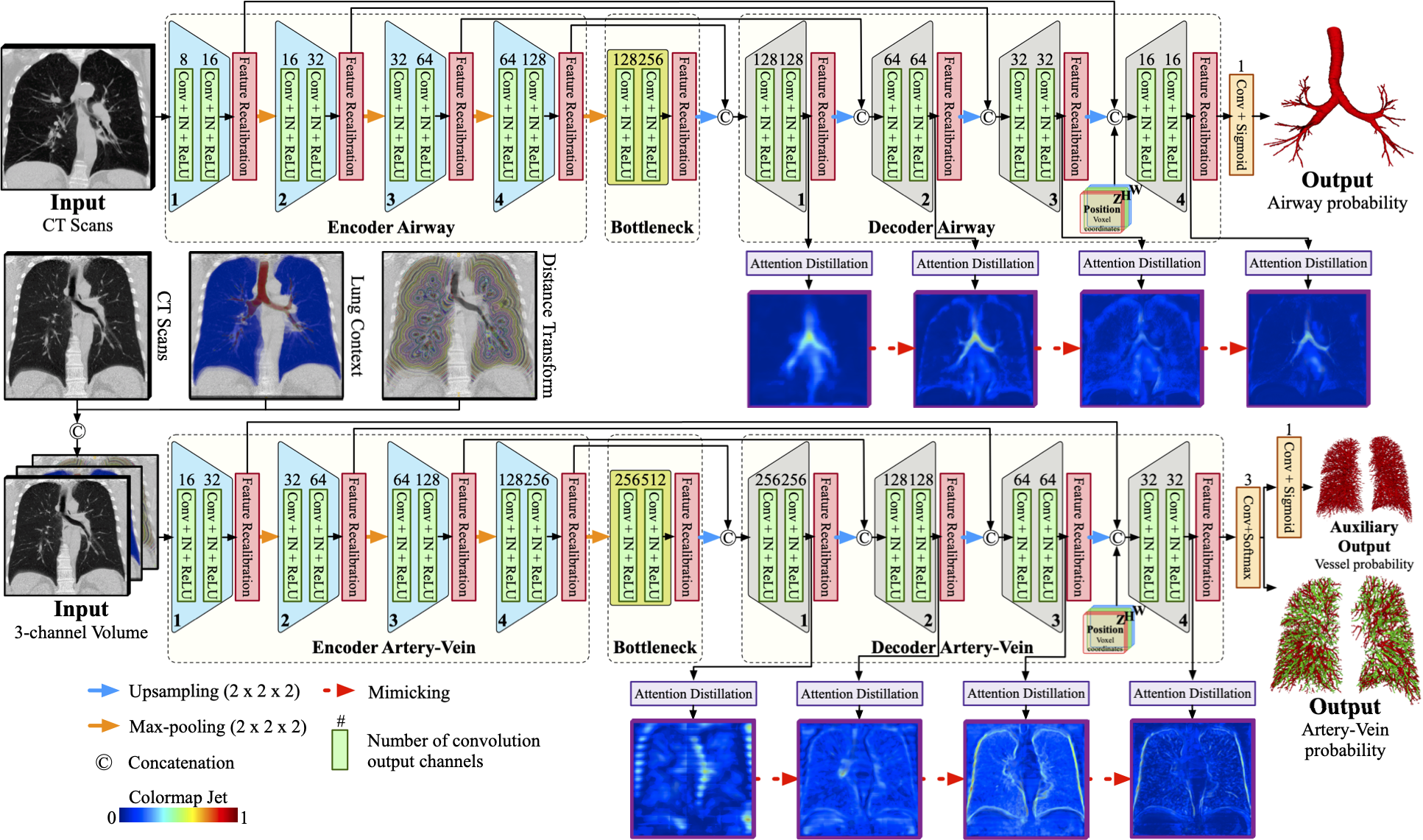}}
\caption{{Overview of the proposed method for pulmonary airway and artery-vein segmentation. Instance normalization and ReLU activation are performed after each convolution layer except the last one. The number of convolution kernels is denoted above each layer.}}
\label{fig:arch}
\end{figure*}

\subsection{Contributions}
To address these concerns, we present a CNNs-based method for pulmonary airway and artery-vein segmentation. Since airways, arteries, and veins are all tubular structures, they are collectively referred to as tubules in the present study. With the carefully designed constituent modules, the proposed method learns to comprehend the contour shape, intensity distribution, and connectivity of bronchi and vessels in a data-driven way. It tackles the challenges of applying CNNs to recognition of long, thin tubules and enjoys high sensitivity to bronchioles, arterioles, and venules.

First, we propose a feature recalibration module to maximally utilize the features learned from CNNs. On one hand, to increase the field-of-view for large context comprehension, {deep architectures with multiple convolution and pooling layers are preferred. Accordingly, the number of learnable parameters increases and then overfitting becomes a problem. On the other hand, if the number of feature channels is simply reduced to avoid over-fitting, it might go to the other extreme where the model fails to learn discriminative features. Therefore, feature recalibration is considered because it intensifies task-related features given a moderate model size.} In the design of recalibration module, we hypothesize that spatial information of features is indispensable for channel-wise recalibration and should be treated differently from position to position and layer to layer. The average pooling used in \cite{rickmann2019project, zhu2019anatomynet} for spatial compression may not well capture the location of airways and vessels in different resolution scales. In contrast, we aim at prioritizing information at key positions with learnable weights, which provides appropriate spatial hints to model inter-channel dependency and thereafter improves recalibration.

Second, we introduce an attention distillation module to reinforce representation learning of tubular airway, artery, and vein. Attention maps of different scales enable us to potentially reveal the morphology and distribution pattern of airways and vessels. Inspired by knowledge distillation \cite{Zagoruyko2017AT, hou2019learning}, we refine the attention maps of lower resolution by mimicking those of higher resolution. Finer attention maps (teacher's role) with richer context can cram coarser ones (student's role) with details about airways, arteries, and veins. The model's ability to recognize delicate branches is ameliorated after recursively focusing on the target anatomy. Dealing with insufficient supervisory signals, the distillation itself acts as an auxiliary learning task that provides extra signals to assist training.


Third, we incorporate anatomy prior into artery-vein segmentation by introducing lung context map and distance transform map. The lung context map, containing automatically segmented airway lumen, airway wall, and lung, explicitly informs the model of semantic knowledge. The distance transform map, computed using extracted airways, records the distance of each voxel to its nearest airway wall.

{Fourth}, the proposed end-to-end method is applicable for both pulmonary airway and artery-vein segmentation. We do not perform independent vessel segmentation beforehand and require no post-refinement on the outputs of CNNs. The sliding window-based segmentation is used and each voxel's coordinates within the thoracic cavity are fed into the model to make up for the loss of position information.

{Finally, although the entire framework is an integrated solution to airway and artery-vein segmentation, its constituting components can be considered for designing solutions to other tasks. The proposed method may also be readily extended by incorporating traditional techniques as post-processing (e.g., graph-cuts \cite{boykov2001fast}), where explicit graph and connectivity modeling are introduced specifically for tubular structures.}

Our contributions can be briefly summarized as follows:
\begin{itemize}
\item We present a tubule-sensitive CNNs-based method for pulmonary airway and artery-vein segmentation. To our best knowledge, this method represents the first attempt to segment airways, arteries, and veins simultaneously.
\item We propose a feature recalibration module that integrates prioritized spatial knowledge for channel-wise recalibration. It encourages discriminative feature learning.
\item We introduce an attention distillation module to reinforce representation learning of tubular airway, artery, and vein. No extra annotation labor is required.
\item We incorporate explicit anatomy prior into artery-vein segmentation by utilizing the lung context map and distance transform map as additional inputs.
\item We respectively validate the proposed method on 110 and 55 non-contrast clinical CT scans for pulmonary airway and artery-vein segmentation. Extensive experiments show that our method achieved superior sensitivity to thin airways, arteries, and veins, with surpassing or competitive overall segmentation performance maintained.
\end{itemize}

\section{Methods}
\label{sec:methods}
Overview of the proposed airway and artery-vein segmentation methods is illustrated in Fig. \ref{fig:arch}. To fulfill effective feature learning of tubular targets, feature recalibration and attention distillation modules are introduced into CNNs. Anatomy prior is included to provide semantic knowledge for artery-vein task.

Given an input CT volume $X$, our segmentation process can be formulated as $P_{Target}=\mathcal{F}(X)$, where $Target$ can be airway, artery, or vein and $P_{Target}$ denotes its corresponding predicted probability. The objective is to learn an end-to-end mapping $\mathcal{F}$ via CNNs to minimize the difference between $P_{Target}$ and its ground-truth label $Y_{Target}$. Assuming CNNs have $M$ convolution layers in total, we denote the activation output of the $m$-th convolution as $A_m\in R^{C_m\times D_m\times H_m\times W_m}$, $1\leq m\leq M$. The number of its channels, depths, heights, and widths are respectively denoted as $C_m$, $D_m$, $H_m$ and $W_m$.

\subsection{Feature Recalibration}

\begin{figure}[htbp]
\centerline{\includegraphics[width=\columnwidth]{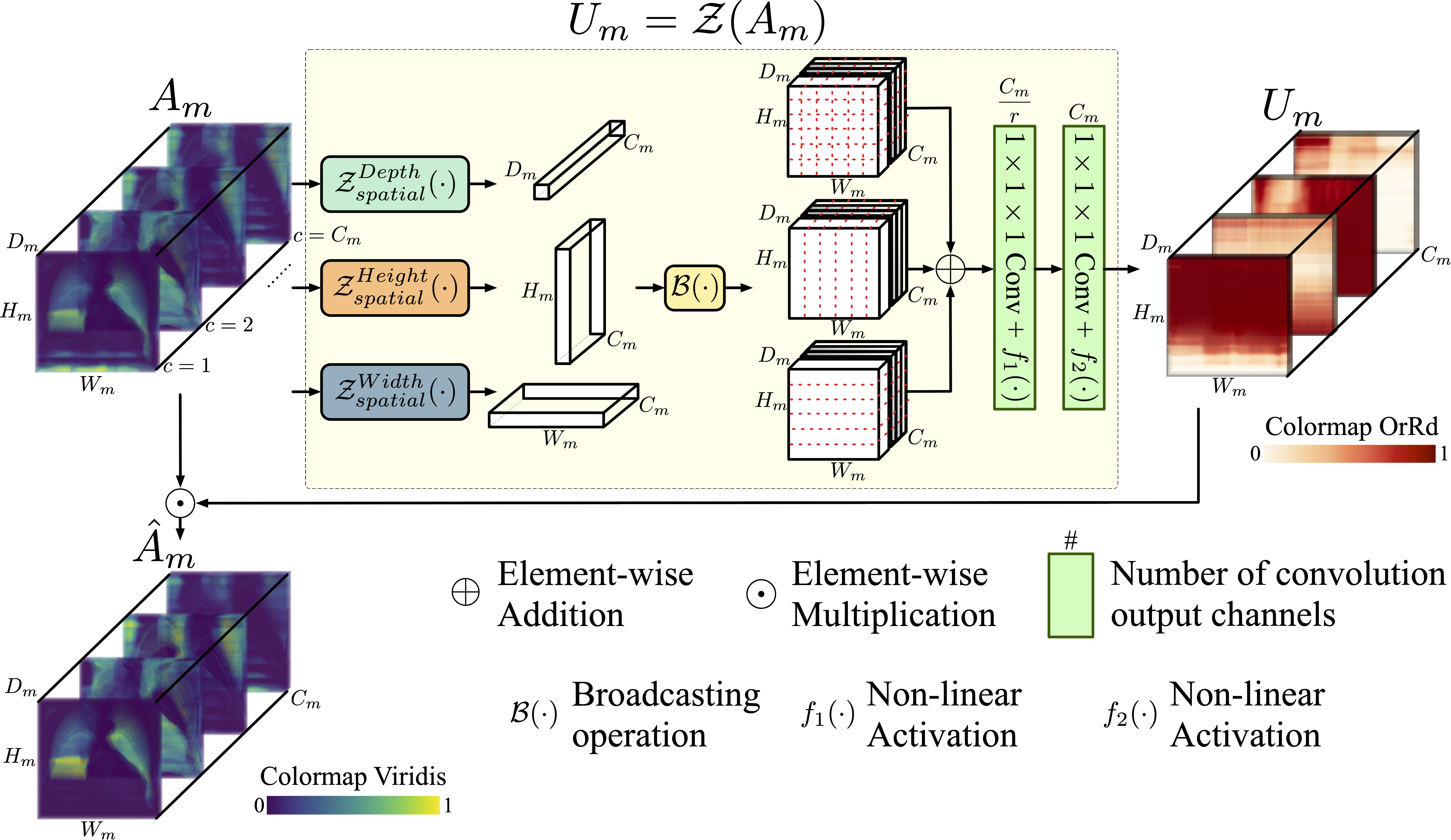}}
\caption{Illustration of the mapping $\mathcal{Z}(\cdot)$ for feature recalibration. Its input is the activated feature $A_m$ of the $m$-th convolution layer. First, spatial map that highlights important regions is integrated through $\mathcal{Z}_{spatial}(\cdot)$ along three axes of depth, height, and width. Second, channel recombination is performed on the spatial map to compute the channel descriptor $U_{m}$. The final element-wise multiplication between $A_m$ and $U_{m}$ produces the recalibrated feature $\hat{A}_{m}$. The notations $r$, $C_m$, $D_m$, $H_m$, and $W_m$ refer to the channel compression factor, the number of channels, depths, heights, and widths of $A_m$, respectively.}
\label{fig:recali}
\end{figure}

We propose the mapping $\mathcal{Z}(\cdot)$ that generates a channel descriptor $U_m = \mathcal{Z}(A_{m})$ to recalibrate the activated convolutional feature $A_{m}$. An overview of $\mathcal{Z}(\cdot)$ for feature recalibration is given in Fig. \ref{fig:recali}. The channel-wise weight map $U_{m}$ is learned to not only unearth crucial spatial locations of $A_{m}$, but also strengthen basis channels that affect most the output decision. The mapping $\mathcal{Z}(\cdot)$ is composed of two steps: 1) spatial knowledge integration to obtain the compressed spatial map; 2) channel-wise recombination of such spatial map. For the first step, previously proposed feature recalibration methods \cite{rickmann2019project, zhu2019anatomynet} treated all positions equally by condensing spatial information of features into vector or scalar value with average pooling, which might not be appropriate for processing large volumetric features. Instead, we integrate spatial information using weighted combination of features along each spatial dimension. Our hypothesis is that different positions may hold different degrees of importance both within the current $A_m$ and across resolution scales (e.g., the shallower $A_{m-1}$ and the deeper $A_{m+1}$). An operation like adaptive or global pooling is not spatially discriminating between the finest features (containing thin bronchioles, arterioles, and venules which are easily ``erased" by averaging) and the coarsest features (containing mostly thick bronchi and vessels). Therefore, we introduce the following spatial integration method $\mathcal{Z}_{spatial}(\cdot)$ that preserves relatively important regions. It can be formulated as:
\begin{equation}
\begin{split}
\mathcal{Z}_{spatial}(A_{m}) &= \mathcal{B}(\mathcal{Z}_{spatial}^{Depth}(A_{m})) + \mathcal{B}(\mathcal{Z}_{spatial}^{Height}(A_{m}))\\ &+ \mathcal{B}(\mathcal{Z}_{spatial}^{Width}(A_{m})),
\end{split}
\label{eq:spatialmapping}
\end{equation}
\begin{equation}
\begin{split}
\mathcal{Z}_{spatial}^{Depth}(A_{m}) &= \sum_{j=1}^{H_m} h_{j}\sum_{k=1}^{W_m} w_{k}A_{m}[:,:,j,k],\\ \mathcal{Z}_{spatial}^{Depth}(A_{m})&\in R^{C_m\times D_m\times 1\times 1},
\label{eq:spatialmappingd}
\end{split}
\end{equation}
\begin{equation}
\begin{split}
\mathcal{Z}_{spatial}^{Height}(A_{m}) &= \sum_{i=1}^{D_m}d_i\sum_{k=1}^{W_m}w_k A_{m}[:,i,:,k],\\ \mathcal{Z}_{spatial}^{Height}(A_{m})&\in R^{C_m\times 1\times H_m\times 1},
\label{eq:spatialmappingh}
\end{split}
\end{equation}
\begin{equation}
\begin{split}
\mathcal{Z}_{spatial}^{Width}(A_{m}) &= \sum_{i=1}^{D_m}d_i\sum_{j=1}^{H_m}h_j A_{m}[:,i,j,:],\\ \mathcal{Z}_{spatial}^{Width}(A_{m})&\in R^{C_m\times 1\times 1\times W_m},
\label{eq:spatialmappingw}
\end{split}
\end{equation}
\noindent where indexed slicing (using Python notation) and broadcasting $\mathcal{B}(\cdot)$ are performed. Notations $C_m$, $D_m$, $H_m$, and $W_m$ are referring to the number of channels, depths, heights, and widths of the $m$-th layer feature $A_m$. The learnable parameters $d_i, h_j, w_k$ denote the combination weights for each feature slice in depth, height, and weight dimension, respectively. During training, crucial airway and artery-vein regions are gradually preferred with higher weights while uninformative corner regions are neglected with lower weights. For the second step, we apply the excitation technique \cite{rickmann2019project} on the compressed spatial map to model inter-channel dependency. Specifically, the channel descriptor $U_m$ is obtained by:
\begin{equation}
U_m = \mathcal{Z}(A_{m})=f_2(K_2*f_1(K_1*\mathcal{Z}_{spatial}(A_{m}))),
\label{eq:channelmapping}
\end{equation}
\noindent where $K_1, K_2$ are 3-D kernels of size 1$\times$1$\times$1 and ``$*$" denotes convolution. Convolving with $K_1$ decreases the channel number to $C_m/r$ and that with $K_2$ recovers back to $C_m$. The ratio $r$ is the compression factor that determines reduction extent. $f_1(\cdot)$ and $f_2(\cdot)$ are non-linear activation functions. We choose Rectified Linear Unit (ReLU) as $f_1(\cdot)$ and Sigmoid as $f_2(\cdot)$ in the present study. Multiple channels are recombined through such channel reduction and increment, with informative ones emphasized and redundant ones suppressed. Given the activated convolutional feature $A_{m}$ and its channel descriptor $U_m$, the recalibrated feature $\hat{A}_{m}$ is defined as:
\begin{equation}
\hat{A}_{m}=U_m\odot A_m,
\label{eq:recalibration}
\end{equation}
\noindent where $\odot$ denotes element-wise multiplication.

\subsection{Attention Distillation}
In both airway and artery-vein segmentation tasks, the segmentation model is required to identify thin tubules like distal bronchi, arteries, and veins. It could be expected that reinforced attention on such objects during feature learning may conduce to improved performance. Recent studies \cite{Zagoruyko2017AT, hou2019learning} on knowledge distillation showed that attention maps serve as valuable knowledge and can be transferred layer-by-layer from teacher networks to student networks. Motivated by knowledge transferability and self-attention mechanism, we introduce the attention distillation module into our 3-D CNNs for recognition of narrow, thin objects. The activation-based attention maps, which guide where to look at, are distilled and exploited during backward transfer process. Without separately setting two different models, later layers play the role of teacher and ``impart" such attention to earlier layers within the same model. Besides, to tackle insufficient supervisory signals caused by the severe class imbalance, the distillation can be viewed as another source of supervision. It produces additional gradients by forcing low-resolution attention maps to resemble high-resolution ones, aiding the training of deep CNNs. Specifically, the attention distillation is performed between two consecutive features $A_m$ and $A_{m+1}$.

\begin{figure}[htbp]
\centerline{\includegraphics[width=\columnwidth]{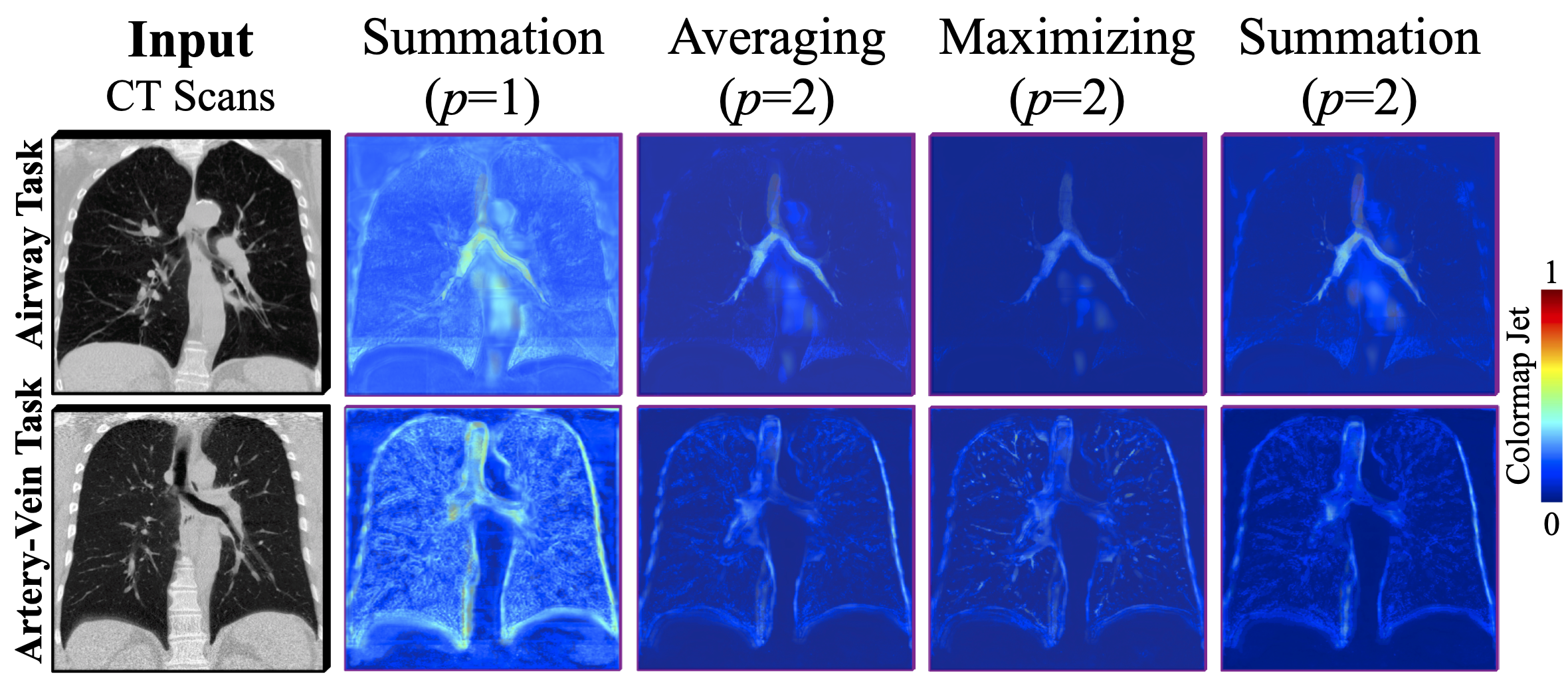}}
\caption{{Difference among mapping functions $\mathcal{G}(\cdot)$ of computing the last attention map in decoder for airway and artery-vein segmentation tasks.}}
\label{fig:attcmp}
\end{figure}

Firstly, the attention map is generated by $G_{m}=\mathcal{G}(A_{m}), G_{m}\in R^{1\times D_m\times H_m\times W_m}$. Each voxel's absolute value in $G_{m}$ reflects the contribution of its correspondence in $A_m$ to the entire segmentation model. One way of constructing the mapping function $\mathcal{G}(\cdot)$ is to compute the statistics of activation values $A_{m}$ across channel:
\begin{equation}
\begin{split}
G_{m}=\sum_{c=1}^{C_m}\lvert A_m[c,:,:,:]\rvert^p,
\end{split}
\label{eq:activation_map}
\end{equation}
The element-wise operation $\lvert \cdot\rvert^{p}$ denotes the absolute value raised to the $p$-th power. More attention is addressed to highly activated regions if $p>1$. Here, we adopt channel-wise summation instead of maximizing $\max_{c}(\cdot)$ or averaging $\frac{1}{C_m}\sum_{c=1}^{C_m}(\cdot)$ because it is relatively less biased. The sum operation retains all implied salient activation information without ignoring non-maximum elements or weakening discriminative elements. For intuitive comparison of different $\mathcal{G}(\cdot)$, visualization of 3-D attention maps on 2-D plane is presented in Fig. \ref{fig:attcmp} by first choosing multiple 2-D slices that contain airways and vessels and then super-imposing them together with opacity of 30\%. Visual comparison exhibits that summation with $p>1$ intensifies most the sensitized task-related regions (e.g., lung borders, bronchi, vessels).

Secondly, trilinear interpolation $\mathcal{I}(\cdot)$ is performed to ensure that processed 3-D attention maps share the same dimension.

Then, voxel-wise Softmax $\mathcal{S}(\cdot)$ is spatially applied to normalize all elements in $[0,1]$. Finally, we drive the distilled attention $\hat{G}_m$ closer to $\hat{G}_{m+1}$ by minimizing the loss:
\begin{equation}
\mathcal{L}_{distill} = \sum_{m=1}^{M-1}\lVert\hat{G}_m-\hat{G}_{m+1}\rVert^2_{F}, \hat{G}_m = \mathcal{S}(\mathcal{I}(G_{m})),\label{eq:distillation}
\end{equation}
\noindent where $\lVert\cdot\rVert^2_{F}$ is the squared Frobenius norm. With $\hat{G}_m$ recursively mimicking its successor $\hat{G}_{m+1}$, visual attention is transmitted from the deepest to the shallowest layer. Note that such distillation process does not require extra annotation labor and can work with arbitrary CNNs readily. In implementation, to prevent the latter attention $\hat{G}_{m+1}$ from approximating the previous $\hat{G}_{m}$, we detach $\hat{G}_{m+1}$ from the computation graph for each $m$ in loss calculation. Consequently, $\hat{G}_{m+1}$ will not be changed by back-propagating errors. The reasons why we do not down-sample $\hat{G}_{m+1}$ to the size of $\hat{G}_{m}$ is that $\hat{G}_{m+1}$ at decoder side has higher resolution than $\hat{G}_{m}$ by nature and down-sampling loses rich information that only exists in $\hat{G}_{m+1}$. It is necessary to keep $\hat{G}_{m+1}$ unchanged so that the resultant distillation loss between $\hat{G}_{m}$ and $\hat{G}_{m+1}$ can improve model's attention on fine details about targets.

\subsection{Anatomy Prior For Artery-Vein Segmentation}
\label{sec:anatomyprior}

\begin{figure}[htbp]
\centerline{\includegraphics[width=\columnwidth]{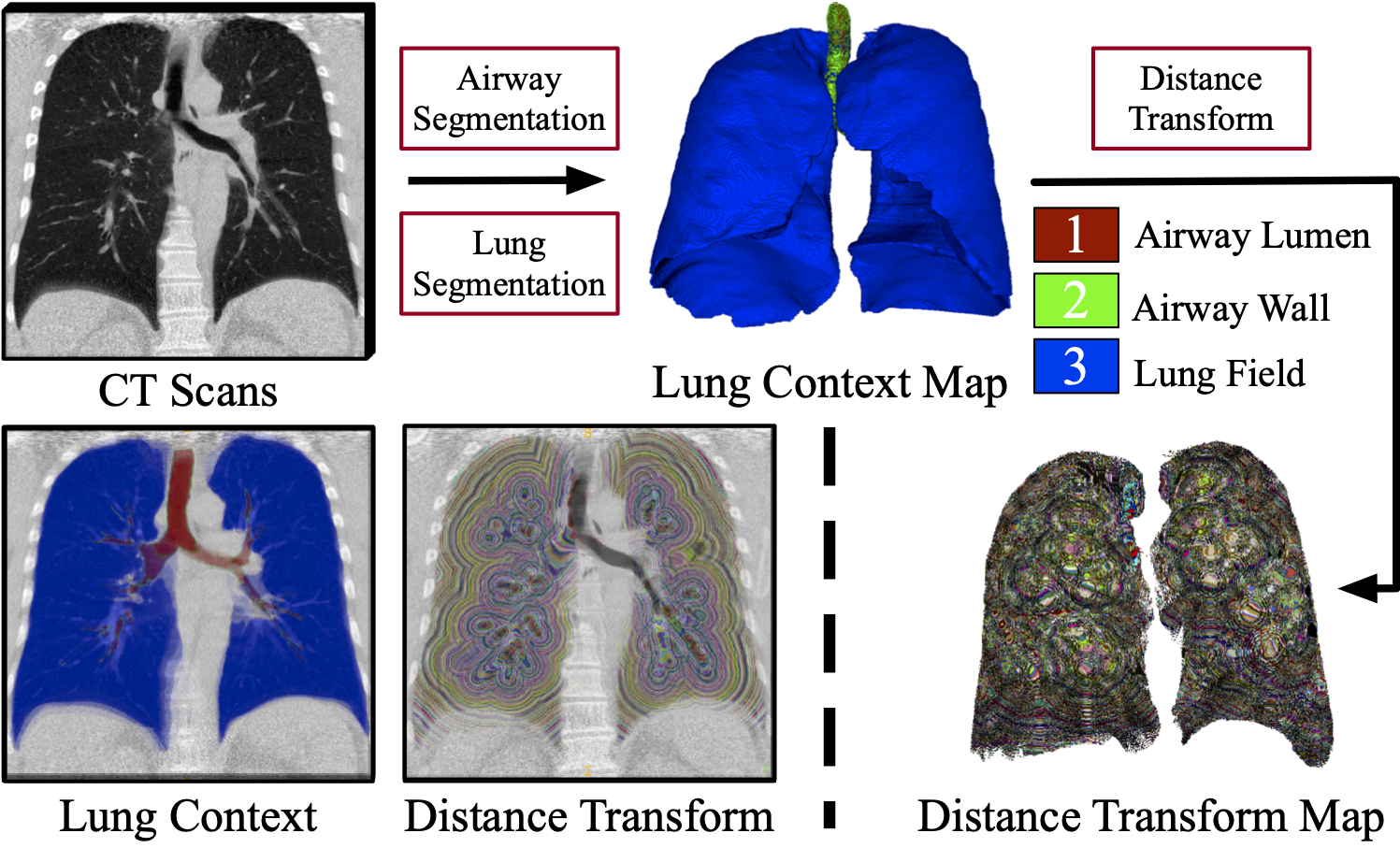}}
\caption{Illustration of anatomy prior incorporation. Visual display of the generated lung context maps and distance transform maps superimposed on CT scans is given in bottom left.}
\label{fig:anatomy}
\end{figure}

In the present study, artery-vein segmentation in the lung hilum (e.g., pulmonary trunk, left and right main pulmonary veins) is excluded since recognition of these vessels in non-contrast CT is extremely difficult for both computers and medical experts \cite{charbonnier2015automatic}. Considering that the valid artery and vein targets are mainly restricted inside the two lungs, we believe it is reasonable to provide segmented lung masks as VOI hint. The lung segmentation is performed by: 1) binarization using OTSU thresholding \cite{otsu1979threshold}; 2) hole filling using morphological operations; 3) selection of the two largest connected components as left and right lungs; 4) convex hull computation to prevent over-segmentation. Besides, we hypothesize that proper representation of airway anatomy is beneficial for the model to distinguish between vessels and airway walls, where similar intensity distribution is shared. Therefore, automatic airway segmentation is first performed using the proposed method to obtain airway lumen. Then, assuming the thickness of airway wall is less than 2 mm \cite{van2010automatic}, we extract airway wall by subtracting airway lumen from its morphological dilation result. The structuring element for dilation is a sphere with diameter of 3 voxels. Given the segmented airway lumen, wall, and lung field, we respectively label them as 1, 2 and 3 to generate the lung context map.

Since pulmonary arteries inside lungs often accompany with airways in parallel, we believe the proximity of arteries to airways might be informative for the segmentation model to discriminate arteries from veins \cite{hislop2002airway, miller1947lung, berend1979relationship, kandathil2018pulmonary}. Consequently, Euclidean distance transform is performed on the segmented airways to calculate the distance of each voxel to its nearest airway wall. The computed distance transform map is multiplied with lung mask to keep valid regions.

To summarize, two maps are introduced as anatomy prior for artery-vein segmentation (see Fig. \ref{fig:anatomy}): lung context map and distance transform map. The first map offers extra semantic knowledge of lung and the second map reflects voxels' closeness to airway. These maps are concatenated with CT sub-volume as inputs to the artery-vein segmentation model.

\subsection{Model Design}
The proposed method employs 3-D U-Net \cite{cciccek20163d} as network backbone. Such encoder-decoder CNNs first extract a condensed representation of input image and then reconstruct it in response to different tasks. To enlarge the receptive field of CNNs and facilitate feature learning of long-range relationship, four pooling layers are used with five resolution scales involved in total. At each scale, both encoders and decoders have two convolution layers (kernel size $3\times3\times3$) followed by instance normalization and ReLU. The feature recalibration module is inserted at the end of each resolution scale. Since high-level features in decoders are also of high-resolution and high-relevance to segmented targets, we perform the decoder-side attention distillation to pass down the fine-grained details that are missing in previous low-resolution attention maps. The encoder-side distillation is not favored because low-level features are more local-scale and general. Furthermore, voxel coordinate map, which records voxels' global position inside the thoracic cavity, is concatenated with features at decoder 4 to make the model explicitly consider location. In view of the patch-wise training, such coordinate map is used to offset the loss of position information. Since both arteries and veins are vessels, we introduce an auxiliary task of vessel segmentation by adding another convolution layer with sigmoid activation to the artery-vein output. Such multi-head design takes advantage of: 1) the inclusion relationship between vessel and artery-vein; 2) the reduced difficulty of learning to recognize vessels. Preliminary ablation study on the auxiliary vessel segmentation output has confirmed its effectiveness. The probability outputs of airway and artery-vein are respectively obtained using sigmoid and softmax activation. Preliminary experiments confirmed that such model design and feature number choice are optimum for our tasks.

\subsection{Training Loss}
To deal with hard samples, we use both the Dice \cite{milletari2016v} and Focal loss \cite{lin2017focal} for training CNNs. For airway segmentation, given the binary label $y^{a}(x)$ and prediction $p^{a}(x)$ of each voxel $x$ in the volume set $X$, the combined loss is defined as:
\begin{equation}
\begin{split}
\mathcal{L}_{Airway} &= - (\frac{2\sum_{x\in X}p^{a}(x)y^{a}(x)}{\sum_{x\in X}(p^{a}(x)+y^{a}(x))+\epsilon}\\ & +\frac{1}{\lvert X\rvert}\sum_{x\in X}(1-p^{a}_{t}(x))^{2}\log(p^{a}_{t}(x))),
\end{split}
\label{eq:airwayloss}
\end{equation}
\noindent where $p^{a}_{t}(x)=p^{a}(x)$ if $y^{a}(x)=1$. Otherwise, $p^{a}_{t}(x)=1-p^{a}(x)$. Parameter $\epsilon$ is used to avoid division by zero. For multi-class artery-vein and binary vessel segmentation tasks, the losses are defined as the following:
\begin{equation}
\begin{split}
\mathcal{L}_{A-V} &= - \frac{1}{3}\sum_{i=0}^{2}(\frac{2\sum_{x\in X}p^{av}_{i}(x)y^{av}_{i}(x)}{\sum_{x\in X}(p^{av}_{i}(x)+y^{av}_{i}(x))+\epsilon}\\ & +\frac{1}{\lvert X\rvert}\sum_{x\in X}(1-p^{av}_{it}(x))^{2}\log(p^{av}_{it}(x))),\\
\mathcal{L}_{Vessel} &= - (\frac{2\sum_{x\in X}p^{v}(x)y^{v}(x)}{\sum_{x\in X}(p^{v}(x)+y^{v}(x))+\epsilon}\\ & +\frac{1}{\lvert X\rvert}\sum_{x\in X}(1-p^{v}_{t}(x))^{2}\log(p^{v}_{t}(x))),
\end{split}
\label{eq:avloss}
\end{equation}
\noindent where $p^{av}_{i}(x)$ and $y^{av}_{i}(x)$ are respectively the artery-vein prediction and label of the $i$-th class ($i=0$ for background, $1$ for artery, and $2$ for vein), $p^{av}_{it}(x)=p^{av}_{i}(x)$ if $y^{av}_{i}(x)=1$ and otherwise $p^{av}_{it}(x)=1-p^{av}_{i}(x)$. The vessel prediction and label are respectively denoted as $p^{v}(x)$ and $y^{v}(x)$, with $y^{v}(x)=1$ if $x$ belongs to artery or vein. Similarly, we have $p^{v}_{t}(x)=p^{v}(x)$ if $y^{v}(x)=1$. Otherwise, $p^{v}_{t}(x)=1-p^{v}(x)$. The total losses for training the airway segmentation model and artery-vein segmentation model are respectively:
\begin{equation}
\mathcal{L}_{Airway}^{total} = \mathcal{L}_{Airway} + \alpha\mathcal{L}_{distill}
\label{eq:lossall1}
\end{equation}
\begin{equation}
\mathcal{L}_{A-V}^{total} = \mathcal{L}_{A-V} + \mathcal{L}_{Vessel} + \alpha\mathcal{L}_{distill},
\label{eq:lossall2}
\end{equation}
where $\alpha$ balances these terms.


\section{Experimental Setup}
\label{sec:experi}

\subsection{Materials}
In total, 110 and 55 non-contrast CT scans from multiple sources were respectively used for airway and artery-vein segmentation. We respectively conducted experiments for these two tasks with non-overlapping datasets. The acquisition and investigation of data were conformed to the principles outlined in the declaration of Helsinki \cite{world2001world}.


\subsubsection{Pulmonary airway task} We used 110 chest CT scans for airway segmentation. They were collected from two public datasets: a) 70 scans from LIDC-IDRI \cite{armato2011lung}; b) 40 scans from the EXACT'09 \cite{lo2012extraction}. The LIDC-IDRI dataset contains 1018 CT scans with pulmonary nodule annotations. In view of image quality, 70 scans with slice spacing less than 0.625 mm were randomly chosen. The EXACT'09 Challenge provides 20 scans in the training set and 20 scans in the testing set. No airway annotation is openly available. The axial size of all CT scans is 512$\times$512 pixels with spatial resolution of 0.5--0.781 mm. The number of slices in each scan varies from 157 to 764. Their slice spacing ranges between 0.45--1.0 mm. The airway ground-truth of each CT scan was acquired by: a) performing an interactive segmentation via ITK-SNAP \cite{py06nimg} to generate a rough airway tree; b) manual delineation and correction by well-trained experts. All 70 scans from LIDC-IDRI and 20 scans of the EXACT'09 training set were used in model training and evaluation. These 90 scans were randomly split into the training set (63 scans), validation set (9 scans) and testing set (18 scans). In addition, all 20 scans of the EXACT'09 testing set were reserved as an independent evaluation set. Segmentation results on this extra dataset were evaluated by EXACT'09 organizers for fair comparison experiments.

\subsubsection{Pulmonary artery-vein task} We used all 55 non-contrast chest CT scans from CARVE14 \cite{charbonnier2015automatic} for artery-vein segmentation. These scans share the same axial size of 512$\times$512 pixels. They have the same slice spacing of 0.7 mm and the spatial resolution is 0.59--0.83 mm. The number of axial slices ranges from 349 to 498. Two kinds of artery-vein reference were available: 1) full annotations of 10 CT scans; 2) partial annotations of a small portion of vessel segments for the remaining 45 CT scans. We randomly split these 45 CT scans into the training set (40 scans) and validation set (5 scans). The 10 CT scans with full artery-vein labels were kept as the testing set. Since the number of labeled vessel segments of the 45 scans was too small to train and validate CNNs, we used semi-automatic segmentation results released by \cite{charbonnier2015automatic} as complement target labels. Due to annotation difficulty, some voxels were marked as non-determined by two observers and we excluded these voxels in training and evaluation. Vessel roots, which are large main vessels entering from the lung hilum into the lung field, were not marked as they are too difficult to delineate in non-contrast CT \cite{charbonnier2015automatic}. Disagreement between observers exists and vessel annotations in this region are unavailable. Therefore, the segmentation targets of interest are limited inside lungs.

\subsection{Implementation Details}
\label{sec:implementdetails}
Since CT scans were acquired from different scanners using different parameter settings, data pre-processing is imperative before model training. The voxel intensity of all scans was truncated within the Hounsfield Unit (HU) window of $[-1000, 400]$ and normalized to $[0, 1]$. To avoid learning irrelevant marginal area outside lung, lung mask was extracted using the method mentioned in Sec. \ref{sec:anatomyprior}. The minimum bounding box of lung was cropped as valid input region. Here, isotropic resampling is not used because it triggers off mismatch between CT images and ground-truth labels during voxel interpolation. The resampled annotations are discontinuous and incomplete, which are detrimental for CNNs to learn effective representation of long, thin, tubular targets. Due to GPU memory limit, CT scans were respectively cropped into sub-volume cubes of the size $80\times192\times304$ and $64\times176\times176$ for airway and artery-vein tasks. {Such size was chosen specifically to maximally utilize the available GPU resource but is not related to the shape of the lung. Given the same GPU memory, the size of cropped cubes for artery-vein task is smaller than that for airway task because the proposed artery-vein model has around 4 times as many parameters as the proposed airway model.} The size of the cropped cubes is kept the same for all phases of training, validation, and testing. In training phase, random horizontal flipping, shifting, Gaussian smoothing ($\sigma=1$), and voxel intensity jittering were applied as on-the-fly data augmentation. The Adam optimizer was used with an initial learning rate of $3\times10^{-3}$. If the training loss stayed at a plateau for over 10 epochs, the learning rate was reduced by a factor of 10. With batch size of 1, training converged after 60 epochs for each model. In validation and testing phases, we performed sliding window prediction with axial stride of 64. Results were averaged on overlapping margins. Each voxel's category of artery-vein or airway was assigned by respectively performing channel-wise $\arg\max$ or thresholding ($th=0.5$) on the probability outputs. No post-processing was involved. All models were implemented in Python with PyTorch or Keras. Model training and hyper-parameter tuning were performed only on the training set. The model that achieved the best validation results was chosen and tested on the testing set for objectivity. In experiments, model training was executed on a Linux workstation with Intel Xeon Silver 4114 CPU, 192 GB RAM, and NVIDIA Tesla V100 GPU. Model inference and anatomy prior computation were carried out on a Linux PC with Intel Core i7-8700 CPU, 64 GB RAM, and NVIDIA Quadro P4000 GPU. The computational time of the proposed airway and artery-vein segmentation method is reported in Table \ref{table:time} under the current hardware configuration. For hyper-parameter settings, we empirically chose $\alpha=0.1$, $\epsilon=10^{-7}$, $p=2$ and $r=2$. Note that current settings worked well for our tasks but are not necessarily optimum. Elaborate tuning may be conducted in specific tasks.

\begin{table}[tbp]
\centering
\caption{Computational time of the proposed pulmonary airway and artery-vein segmentation method.}
\label{table:time}
\scalebox{0.76}{
\begin{tabular}{l|l|l}
\hline
\multicolumn{2}{l|}{Item Name}                                                                                                                                                                    & Time (seconds) \\ \hline
\multirow{2}{*}{\begin{tabular}[c]{@{}l@{}}Pulmonary Airway\\ Segmentation\end{tabular}}      &  \begin{tabular}[c]{@{}l@{}}Training\\ (\# Epoch$\times$\\ Time Per Epoch)\end{tabular} & \begin{tabular}[c]{@{}l@{}}60$\times$\\ 3294.0$\pm$31.1\end{tabular} \\ \cline{2-3}
& \begin{tabular}[c]{@{}l@{}}Inference\\ (Per CT Volume)\end{tabular}                               &  48.4$\pm$18.5   \\ \hline
\multirow{4}{*}{\begin{tabular}[c]{@{}l@{}}Anatomy Prior\\ (Per CT Volume)\end{tabular}} & Lung Segmentation & 115.9$\pm$95.8  \\ \cline{2-3}
 & \begin{tabular}[c]{@{}l@{}}Airway Segmentation\\ (Inference)\end{tabular} &  45.4$\pm$8.9  \\ \cline{2-3} 
 & Lung Context Map &  3.7$\pm$0.4 \\ \cline{2-3} 
 & Distance Transform Map &  9.2$\pm$1.1 \\ \hline
\multirow{2}{*}{\begin{tabular}[c]{@{}l@{}}Pulmonary Artery-Vein\\ Segmentation\end{tabular}} & \begin{tabular}[c]{@{}l@{}}Training\\ (\# Epoch$\times$\\ Time Per Epoch)\end{tabular}  & \begin{tabular}[c]{@{}l@{}}  60$\times$\\ 2856.7$\pm$153.9 \end{tabular} \\ \cline{2-3} 
& \begin{tabular}[c]{@{}l@{}}Inference\\ (Per CT Volume)\end{tabular}                               &  75.9$\pm$13.9     \\ \hline
\end{tabular}
}
\end{table}

\subsection{Evaluation Metrics}
For airway task, only the largest connected component of the binarized segmentation output was kept in view of clinical practice. Five metrics were used: (a) Branches detected (BD); (b) Tree-length detected (TD); (c) True positive rate (TPR); (d) False positive rate (FPR); (e) Dice similarity coefficient (DSC). We referred to \cite{lo2012extraction} for definitions of metrics (a)--(d). The first two metrics are centerline-based measurements. We computed the centerlines of reference annotations using the algorithm described in \cite{lee1994building}. Then, the centerlines were multiplied with segmentation results to compute the length of the overlapped centerlines $L_{seg}$. The fraction of the correctly segmented tree's length relative to the total tree length of the reference centerlines $L_{ref}$ is defined as TD, $TD=\frac{L_{seg}}{L_{ref}}\times100\%$. For any branch segment between two nodes (bifurcation node or terminal node) on the reference centerlines, if the segmentation results and this segment overlap with over 1 voxel, then this branch is counted as ``detected". The number of branches that are successfully detected $N_{seg}$ with respect to the total number of branches in reference $N_{ref}$ is defined as BD, $BD=\frac{N_{seg}}{N_{ref}}\times100\%$. The metrics of TPR, FPR, and DSC are voxel-based measurements. TPR is defined as the number of true airway voxels in segmentation results $N_{TP}$ divided by the total number of airway voxels in reference $N_{P}$, $TPR=\frac{N_{TP}}{N_{P}}\times100\%$. FPR is defined as the number of false airway voxels in segmentation results $N_{FP}$ divided by the total number of background voxels in reference $N_{N}$, $FPR=\frac{N_{FP}}{N_{N}}\times100\%$. With the categorized $N_{TP}$, $N_{FP}$, and $N_{P}$, DSC is given by: $DSC=\frac{2\times N_{TP}}{N_{TP}+N_{FP}+N_{P}}\times100\%$. Note that trachea region is excluded in calculating BD, TD, TPR, and FPR to reflect model's ability to extract peripheral airways. However, trachea is included in DSC computation as it measures overall segmentation quality.

For artery-vein task, six metrics were used: (a) Accuracy (ACC); (b) TPR; (c) FPR; (d) DSC; (e) BD; (f) TD. All connected components of artery and vein subtrees were involved in measurements. We followed \cite{charbonnier2015automatic} to report both mean and median ACC of artery-vein separation, with 95\% confidence interval (CI) estimated. Other metrics were reported in mean $\pm$ standard deviation. The definitions of BD and TD are the same as those in airway tasks except that arteries and veins are first measured respectively to obtain the number of detected branches ($N_{seg}^{artery}$, $N_{seg}^{vein}$) and the total length of segmented subtrees ($L_{seg}^{artery}$, $L_{seg}^{vein}$). Then, BD and TD are given as the averaged artery and vein results over their corresponding ground-truth: $BD=\frac{1}{2}\times(\frac{N_{seg}^{artery}}{N_{ref}^{artery}}+\frac{N_{seg}^{vein}}{N_{ref}^{vein}})\times100\%$, $TD=\frac{1}{2}\times(\frac{L_{seg}^{artery}}{L_{ref}^{artery}}+\frac{L_{seg}^{vein}}{L_{ref}^{vein}})\times100\%$.

\section{Results}
\label{sec:results}
Evaluation of the proposed method is structured as follows. First, we provide quantitative results of pulmonary airway and artery-vein segmentation in comparison with state-of-the-art methods. Second, ablation study is conducted to validate each constituting component of our method. Third, qualitative segmentation results are presented for visual analysis.

\subsection{Comparison with State-Of-The-Art Methods}
\subsubsection{Pulmonary Airway Segmentation}\label{cmpaw} Table \ref{table:compaw} reports comparison results with state-of-the-art pulmonary airway segmentation methods. Since we adopted U-Net as network backbone, comparison experiments were performed with other encoder-decoder CNNs: the original 3-D U-Net \cite{cciccek20163d}, its variants V-Net \cite{milletari2016v}, VoxResNet \cite{chen2018voxresnet}, and Attention-Gated (AG) U-Net \cite{schlemper2019attention}. The network architecture of these methods has similar encoding path but varied decoding path. We also compared our method with five state-of-the-art methods: Wang et al. \cite{wang2019tubular}, Juarez et al. \cite{juarez2019joint}, Qin et al. \cite{qin2019airwaynet}, Juarez et al. \cite{juarez2018automatic} and Jin et al. \cite{jin20173d}. These methods were re-implemented by ourselves and fine-tuned on the same dataset. Only methods in \cite{wang2019tubular, qin2019airwaynet, juarez2018automatic, jin20173d} were reproduced with Keras. Other pulmonary airway or artery-vein segmentation methods were implemented with PyTorch. Furthermore, we evaluated our method on the independent testing set of EXACT'09 \cite{lo2012extraction}. These 20 testing cases were not used for training or fine-tuning. For a fair comparison, results of three available metrics (BD, TD, and FPR) were given by EXACT'09 organizers and shown in Table \ref{table:compex09}.

\begin{table*}[htbp]
\centering
\caption{{Comparison of pulmonary airway segmentation results. The results both under the same binarization threshold and under the same FPR are presented for each method. The FPR is controlled to be the same with 3-D U-Net (under threshold of 0.5) by respectively adjusting the binarization threshold on the probability outputs of each method.}
}
\label{table:compaw}
\scalebox{0.76}{
\begin{threeparttable}
\begin{tabular}{l|l|l|l|l|l|l|l|l|l|l|l|l|l}
\hline
Method    &  Params ($\times10^4$)    & $th$  & BD (\%) & TD  (\%) & TPR (\%) & FPR (\%) & DSC (\%)  & $th$  & BD (\%) & TD  (\%) & TPR (\%) & FPR (\%) & DSC (\%)  \\ \hline
3-D U-Net \cite{cciccek20163d}\tnote{*}  &  477.1 & \multirow{10}{*}{0.5}  & 87.2$\pm$13.7 &  73.8$\pm$18.7     &  85.3$\pm$10.4   & 0.021$\pm$0.015   &  91.5$\pm$2.9   &  0.5    & 87.2$\pm$13.7 &  73.8$\pm$18.7     &  85.3$\pm$10.4   & 0.021$\pm$0.015   &  91.5$\pm$2.9 \\ \cline{1-2}\cline{4-14}
V-Net \cite{milletari2016v} &   1047.1  &  & 91.0$\pm$16.2   &  81.6$\pm$19.5  & 87.1$\pm$13.6    & 0.024$\pm$0.017     & 92.1$\pm$3.6  &  0.6   & 88.2$\pm$21.3   &  79.1$\pm$21.6  & 85.4$\pm$14.9    & 0.021$\pm$0.014     & 92.0$\pm$3.9 \\ \cline{1-2}\cline{4-14} 
VoxResNet \cite{chen2018voxresnet} &  170.9  &   & 88.2$\pm$12.6   & 76.4$\pm$13.7   & 84.3$\pm$10.4    &0.012$\pm$0.009    & 92.7$\pm$3.0  &  0.001   & 91.6$\pm$10.4   & 81.6$\pm$11.2   & 88.3$\pm$8.6    &0.021$\pm$0.012    & 92.9$\pm$2.3  \\ \cline{1-2}\cline{4-14}
{AG U-Net \cite{schlemper2019attention}}  &     621.3     &     &  {93.8$\pm$7.9}         &     {88.2$\pm$9.4}     &   {91.7$\pm$6.6}     &     {0.031$\pm$0.015}   &      {92.5$\pm$2.0}   &     {0.7}      &  {90.7$\pm$12.1}    &   {83.1$\pm$15.8}   &  {87.3$\pm$11.9}   &  {0.021$\pm$0.012}   &  {92.7$\pm$2.6}    \\ \cline{1-2}\cline{4-14}
Wang et al. \cite{wang2019tubular}  &  549.7  &  & 93.4$\pm$8.0   & 85.6$\pm$9.9 & 88.6$\pm$8.8  & 0.018$\pm$0.012    & 93.5$\pm$2.2  &  0.2  & 94.1$\pm$7.7   & 86.7$\pm$9.5  &  89.5$\pm$8.4  & 0.021$\pm$0.012    & 93.3$\pm$2.2    \\ \cline{1-2}\cline{4-14}
Juarez et al. \cite{juarez2019joint} &  5.3     &    & 77.5$\pm$20.9   & 66.0$\pm$20.4   & 77.5$\pm$15.5     & \textbf{0.009$\pm$0.009}   & 87.5$\pm$13.2   &   0.00001  & 86.5$\pm$16.3   & 76.3$\pm$17.8   &   84.7$\pm$12.5     & 0.021$\pm$0.012   & 89.7$\pm$9.2    \\ \cline{1-2}\cline{4-14}
Qin et al. \cite{qin2019airwaynet} &   106.8   &    & 91.6$\pm$8.3   & 82.1$\pm$10.9   & 87.2$\pm$8.9    & 0.014$\pm$0.009    & \textbf{93.7$\pm$1.9}    &  0.005  & 93.7$\pm$6.3   & 85.7$\pm$9.5   & 89.6$\pm$8.0    & 0.021$\pm$0.011    & 93.3$\pm$1.7   \\ \cline{1-2}\cline{4-14}
Juarez et al. \cite{juarez2018automatic} &  352.7   &    & 91.9$\pm$9.2   & 80.7$\pm$11.3   & 86.7$\pm$9.1    & 0.014$\pm$0.009    & 93.6$\pm$2.2    &  0.05   & 93.7$\pm$7.8   & 83.4$\pm$10.1   & 89.3$\pm$7.9    & 0.021$\pm$0.011    & 93.4$\pm$1.9     \\ \cline{1-2}\cline{4-14}
Jin et al. \cite{jin20173d}  & 473.4   &   & 93.1$\pm$7.9   & 84.8$\pm$9.9   &  88.1$\pm$8.5   & 0.017$\pm$0.010    & 93.6$\pm$2.0  &   0.005   & 94.3$\pm$7.3   & \textbf{87.3$\pm$9.1}   &  89.6$\pm$8.0   & 0.021$\pm$0.012    & 93.5$\pm$1.9    \\ \cline{1-2}\cline{4-14}
Our proposed  & 423.1  &   & \textbf{96.2$\pm$5.8} & \textbf{90.7$\pm$6.9}   & \textbf{93.6$\pm$5.0}    &  0.035$\pm$0.014   &  92.5$\pm$2.0    &   0.77   &  \textbf{94.3$\pm$6.6}      &    86.7$\pm$8.5   & \textbf{90.6$\pm$6.7}    &  \textbf{0.021$\pm$0.011}   &  \textbf{93.5$\pm$1.6}      \\ \hline
\end{tabular}
\begin{tablenotes}
\item[*] Feature channels were halved to have similar number of parameters with the proposed method.
\end{tablenotes}
\end{threeparttable}
}
\end{table*}

Table \ref{table:compaw} shows that under the same thresholding value ($th=0.5$), the proposed method achieved the highest BD of 96.2\%, TD of 90.7\%, and TPR of 93.6\% with a compelling DSC of 92.5\%. Such high sensitivity was accompanied with an inferior FPR of 0.035\%. Since the threshold $th$ directly affects airway segmentation results, we adjusted $th$ to enforce the same FPR for all methods. The FPR of 3-D U-Net \cite{cciccek20163d} under $th=0.5$ was chosen as the ``anchor" FPR for alignment. Except V-Net \cite{milletari2016v}, {AG U-Net \cite{schlemper2019attention}}, and the proposed method, all methods have to be thresholded with a rather low $th<0.5$ to control FPR. Under the same FPR, results of state-of-the-art methods are closer to each other than those under the same $th=0.5$. In that case, the proposed method still achieved the highest BD of 94.3\%, TPR of 90.6\%, and DSC of 93.5\% with a competitive TD of 86.7\%.

\begin{table}[htbp]
\centering
\caption{Evaluation results on the EXACT'09 testing set.}
\label{table:compex09}
\scalebox{0.76}{
\begin{threeparttable}
\begin{tabular}{l|l|l|l}
\hline
Participants\tnote{\dag}  & BD (\%) & TD  (\%) & FPR (\%)  \\ \hline
Neko & 35.5$\pm$8.2  & 30.4$\pm$7.4 & 0.89$\pm$1.78 \\ \hline
UCCTeam & 41.6$\pm$9.0  & 36.5$\pm$7.6 & \textbf{0.71$\pm$1.67} \\ \hline
FF\_ITC & 79.6$\pm$13.5  & \textbf{79.9$\pm$12.1} & 11.92$\pm$13.16 \\ \hline
HybAir & 51.1$\pm$10.9  & 43.9$\pm$9.6 & 6.78$\pm$26.60 \\ \hline
MISLAB & 42.9$\pm$9.6  & 37.5$\pm$7.1 & 0.89$\pm$1.64 \\ \hline
NTNU & 31.3$\pm$10.4  & 27.4$\pm$9.6 & 3.60$\pm$3.37 \\ \hline
Our proposed ($th=0.1$) & \textbf{82.0$\pm$9.9} & 79.4$\pm$10.0  & 9.71$\pm$5.59   \\ \hline
Our proposed ($th=0.5$) & 76.7$\pm$11.5 & 72.7$\pm$11.6   & 3.65$\pm$2.86   \\ \hline
Our proposed ($th=0.8$) & 68.8$\pm$13.4  &  62.6$\pm$12.7   &    1.28$\pm$1.29   \\ \hline
\end{tabular}
\begin{tablenotes}
\item[\dag] Results of recent participants were directly quoted here.  
\end{tablenotes}
\end{threeparttable}
}
\end{table}

In Table \ref{table:compex09}, results of recent participants are reported by EXACT'09 organizers and are not publicly accessible. It is impossible to control all FPRs to be the same. Instead, we binarized our probability results with three thresholding values ($th=0.1, 0.5, 0.8$) and submitted them for official evaluation. Different FPR levels are presented as reference. Under $th=0.1$, the proposed method (FPR: 9.71\%) achieved a 2.4\% higher BD and a comparable TD with respect to team FF\_ITC (FPR: 11.92\%). Under $th=0.5$, compared with teams HybAir (FPR: 6.78\%) and NTNU (FPR: 3.60\%), our method (FPR: 3.65\%) achieved an over 25\% higher BD and an over 28\% higher TD. Under $th=0.8$, we (FPR: 1.28\%) obtained over 1.6 times higher BD and TD than teams Neko (FPR: 0.89\%), UCCTeam (FPR: 0.71\%), and MISLAB (FPR: 0.89\%).

\subsubsection{Pulmonary Artery-Vein Segmentation}\label{cmpav} Table \ref{table:compav} gives comparison results with state-of-the-art pulmonary artery-vein segmentation methods. Apart from the well-known medical image segmentation models mentioned in Sec. \ref{cmpaw}, two recently proposed artery-vein classification methods were evaluated: Charbonnier et al. \cite{charbonnier2015automatic} and Nardelli et al. \cite{nardelli2018pulmonary}. Both two methods were developed to recognize arteries and veins from the already segmented vessels, where vessel segmentation was performed independently in advance. The comparison of the proposed method against labels yielded a mean ACC of 90.3\%, a medium ACC of 90.9\%, a TPR of 90.3\%, a FPR of 0.151\%, a DSC of 82.4\%, a BD of 85.4\%, and a TD of 90.9\%. It outperformed state-of-the-art segmentation CNNs by a large margin in ACC, TPR, BD, and TD with comparable FPR and DSC. Admittedly, compared with methods that adopted graph-based representation for artery-vein separation \cite{charbonnier2015automatic, nardelli2018pulmonary}, the proposed method has room for improvement.

\begin{table*}[htbp]
\centering
\caption{{Comparison of pulmonary artery-vein segmentation results.}}
\label{table:compav}
\scalebox{0.76}{
\begin{threeparttable}
\begin{tabular}{l|l|l|l|l|l|l|l|l}
\hline
Method   & Params ($\times10^4$)      & ACC-mean {[}95\%-CI{]} (\%)  &   ACC-median {[}95\%-CI{]} (\%)  & TPR  (\%) & FPR (\%) & DSC (\%)  & BD (\%)  &  TD (\%)  \\ \hline
3-D U-Net \cite{cciccek20163d}\tnote{*} & 1907.4 & 88.3 [85.4,91.1]  & 88.7 [85.3,93.6]  & 88.2$\pm$3.9  & 0.117$\pm$0.043  &      83.8$\pm$2.9      &  82.8$\pm$6.6  &  89.1$\pm$4.5 \\ \hline
V-Net \cite{milletari2016v}\tnote{*} & 1047.1 & 84.4 [80.7,88.1]  &  84.9 [80.3,90.6]   &  84.4$\pm$4.9   &   0.104$\pm$0.047     &  82.7$\pm$4.5   & 80.0$\pm$7.5   &   85.9$\pm$5.9   \\ \hline 
VoxResNet \cite{chen2018voxresnet}\tnote{*} & 170.9 &  87.0 [83.9,90.0] &  86.0 [84.3,93.7]   &  87.2$\pm$4.1   & 0.116$\pm$0.054   &  83.2$\pm$3.4   &  82.4$\pm$8.0   &   88.7$\pm$5.0  \\ \hline
{AG U-Net \cite{schlemper2019attention}}     & 621.4 & {84.6 [80.8,88.5]}  &  {85.9 [80.9,90.3]}   &  {84.7$\pm$5.2}   & {0.098$\pm$0.037}   &  {83.2$\pm$3.7}  &  {77.3$\pm$7.5}  &  {85.4$\pm$6.1}  \\ \hline
Charbonnier et al. \cite{charbonnier2015automatic}\tnote{\dag}  & -- & 92 [88,95]  &  94 [84,96]    &  --   & --   &  --   & --  & -- \\ \hline
Nardelli et al. \cite{nardelli2018pulmonary}\tnote{\dag}   & 504.3 & 94 [91, 96]  &  95 [93,97]    &  --   & --   &  --   & --  & --  \\ \hline
Our proposed & \multirow{3}{*}{1691.0} &  90.3 [87.7,92.9] &  90.9 [87.4,94.6]  &  90.3$\pm$3.5  &   0.151$\pm$0.043  &   82.4$\pm$3.0  &   85.4$\pm$5.3     &      90.9$\pm$3.8        \\ \cline{1-1} \cline{3-9}
{Our proposed + Graph-cuts (a)\tnote{\ddag}}  &    & {92.3 [90.1,94.4]}  &  {92.7 [89.9,95.9]}  & {92.2$\pm$2.9}     &  {0.141$\pm$0.043}  &  {84.2$\pm$2.7}  &  {87.3$\pm$4.8}  &  {92.9$\pm$3.1}     \\ \cline{1-1} \cline{3-9}
{Our proposed + Graph-cuts (b)\tnote{\ddag}}  &    &  {\textbf{97.2 [96.2,98.2]}}  &  {\textbf{97.5 [96.8,98.6]}}  &     {\textbf{97.1$\pm$1.4}}      &  {\textbf{0.015$\pm$0.008}}      &     {\textbf{97.2$\pm$1.3}}  &  {\textbf{95.6$\pm$1.9}}     &     {\textbf{96.8$\pm$1.5}}     \\ \hline
\end{tabular}
\begin{tablenotes}
\item[*] The same auxiliary vessel segmentation task was introduced as the proposed method.
\item[\dag] Results on the same testing set were directly quoted here.
\item[\ddag] {The graph-cuts post-processing was introduced to classify the segmented vessels into arteries and veins. Two possible ways of building vessel graphs are considered: (a) combining both the predicted arteries and veins from the proposed method as vessels; (b) combining both the ground-truth arteries and veins from labels as vessels. Each vessel voxel is regarded as a non-terminal node and is connected to its neighbor vessel nodes, source node (artery), and sink node (vein). For the regional term of graph-cuts, the CNNs' predicted probabilities of being background ($p_0$), artery ($p_1$) and vein ($p_2$) are re-normalized for each vessel voxel: $p_{1}'={p_1}/{(p_1 + p_2)}$ and $p_{2}'={p_2}/{(p_1 + p_2)}$, where $p_0 + p_1 + p_2 = 1$. The two probabilities are respectively used as the weights of edges between each vessel node and the source node or sink node: $w_{source}=p_{1}'$ and $w_{sink}=p_{2}'$. For the boundary term of graph-cuts, the weight of edge between two connected vessel nodes is calculated by the Gaussian kernel of their CT intensity difference: $w_{AB}=\kappa*exp{(-{(I_{A}-I_{B})^2}/{\sigma}})$, where $I_{A}$ and $I_{B}$ are respectively the intensity of vessel node $A$ and node $B$. The $\kappa$ balances between the regional term and boundary term. The $\sigma$ determines how fast the values decay towards zero with increasing intensity dissimilarity. The final cut is obtained by min-cut/max-flow algorithm \cite{boykov2001fast}, assigning each vessel node to artery or vein. In the present study, hyper-parameters $\kappa=8$ and $\sigma=100$ were found as optimal by means of grid search.}
\end{tablenotes}
\end{threeparttable}
}
\end{table*}

\subsection{Ablation Study}
We investigated the validity of key constituents of the proposed method: 1) feature recalibration (FR); 2) attention distillation (AD); 3) anatomy prior (AP). FR and AD were employed in both airway and artery-vein segmentation while AP was only used in artery-vein task. The model trained without FR, AD, and AP was indicated as baseline. Two very recently proposed feature recalibration modules (cSE \cite{zhu2019anatomynet} and PE \cite{rickmann2019project}) were introduced into our baseline for comparison. They were both adapted from the 2-D squeeze-and-excitation \cite{hu2018squeeze} technique for 3-D channel-wise feature recalibration. We replaced all FR with these two modules and trained models from scratch. For assessing AD, deep supervision (DS) \cite{zhu2017deeply} was introduced for comparison. DS allows features of lower resolution to be supervised directly by targets. Specifically, we respectively added one convolution layer (kernel size $3\times3\times3$) and one trilinear upsampling layer to features of decoder 1--3. After sigmoid or softmax activation, these outputs were involved in loss computation for airway or artery-vein segmentation. The key difference between AD and DS is that DS uses segmentation targets as ``hard" supervision. In contrast, AD acts as ``soft" supervision. It guides preceding layers to pay attention to ``hot areas" in latter layers, where voxel-wise supervision of segmentation is not enforced. To evaluate AP, we calculated AP using airway prediction results from 3-D U-Net \cite{cciccek20163d}, V-Net \cite{milletari2016v}, VoxResNet \cite{chen2018voxresnet}, AG U-Net \cite{schlemper2019attention}, and the proposed airway segmentation method. AP methods with different airway sources are respectively referred to as AP (3-D U-Net \cite{cciccek20163d}), AP (V-Net \cite{milletari2016v}), AP (VoxResNet \cite{chen2018voxresnet}), AP (AG U-Net \cite{schlemper2019attention}), and AP (proposed).

\subsubsection{Feature Recalibration} Table \ref{table:abaw} shows that under the same threshold $th=0.5$, all three recalibration modules (cSE \cite{zhu2019anatomynet}, PE \cite{rickmann2019project}, and FR) bring performance gains to baseline in BD, TD, and TPR. Specifically, the proposed FR leads to the highest increase of 4.5\% in BD, 9.5\% in TD, and 5.7\% in TPR. Meanwhile, all these modules more or less worsen FPR and DSC. Under the same FPR, results of different methods become closer than those under $th=0.5$. All methods except baseline are binarized with $th>0.5$ to reduce FPR. Although FPR of baseline is relaxed to be higher, it only achieved the same BD, 0.3\% higher TD, 1.1\% higher TPR, and 0.8\% lower DSC. The proposed FR boosted performance to a BD of 94.2\%, a TD of 87.5\%, a TPR of 90.1\%, and a DSC of 93.2\%. FR outperformed cSE \cite{zhu2019anatomynet} and PE \cite{rickmann2019project} in BD, TD, TPR, and DSC, which is in line with results under $th=0.5$.

Table \ref{table:abav} reveals that compared with baseline (AP), all recalibration modules increase mean ACC, TPR by over 0.7\%, and DSC by over 0.3\%. The baseline with FR obtained a mean ACC of 89.4\%, a median ACC of 90.2\%, a TPR of 89.4\%, a FPR of 0.150\%, a DSC of 82.0\%, a BD of 83.8\%, and a TD of 89.9\%. Both PE \cite{rickmann2019project} and FR share similar results, surpassing cSE \cite{zhu2019anatomynet} in all metrics but FPR and DSC.

\begin{table*}[htbp]
\centering
\caption{Results of ablation study on pulmonary airway segmentation. The results both under the same binarization threshold and under the same FPR are presented for each method. The FPR is controlled to be the same with 3-D U-Net (under threshold of 0.5) by respectively adjusting the binarization threshold on the probability outputs of each method.\protect\\ cSE = channel-Squeeze-Excitation, PE = Project-Excitation, FR = Feature Recalibration,\protect\\ AD = Attention Distillation, DS = Deep Supervision
}
\label{table:abaw}
\scalebox{0.76}{
\begin{tabular}{l|l|l|l|l|l|l|l|l|l|l|l|l|l}
\hline
Method    & Params ($\times10^4$)   & $th$   & BD (\%) & TD (\%) & TPR (\%) & FPR (\%) & DSC (\%) &  $th$   & BD (\%) & TD (\%) & TPR (\%) & FPR (\%) & DSC (\%)   \\ \hline
Baseline &  411.8   & \multirow{8}{*}{0.5}    & 91.6$\pm$9.2   & 81.3$\pm$11.5   &  87.2$\pm$8.6   &  \textbf{0.014$\pm$0.008}    &  \textbf{93.7$\pm$1.7}   &  0.001   & 91.6$\pm$10.4   & 81.6$\pm$11.2   &  88.3$\pm$8.6   &  0.021$\pm$0.012    &  92.9$\pm$2.3    \\ \cline{1-2}\cline{4-14}
+ cSE \cite{zhu2019anatomynet} &   422.8 &   & 95.1$\pm$6.2    &  88.5$\pm$8.3    &    92.4$\pm$5.5    &   0.033$\pm$0.015      &  92.3$\pm$1.9  &  0.83   &   92.1$\pm$7.9    &  83.0$\pm$10.3    &    89.2$\pm$6.9    &   0.021$\pm$0.012      &  92.8$\pm$1.7       \\ \cline{1-2}\cline{4-14}
+ PE \cite{rickmann2019project} &   422.8  &   & 95.7$\pm$5.1   & 88.4$\pm$7.9   &  92.3$\pm$5.9   & 0.037$\pm$0.019    & 91.8$\pm$2.8    &   0.71        & 90.6$\pm$8.9   & 81.5$\pm$11.4   &  86.6$\pm$8.8   & 0.021$\pm$0.013    & 92.1$\pm$2.1    \\ \cline{1-2}\cline{4-14}
+ FR & 423.1  &  & 96.1$\pm$5.9   &  \textbf{90.8$\pm$7.5}   &  92.9$\pm$5.9   & 0.034$\pm$0.016    & 92.3$\pm$2.3    &   0.76  &    94.2$\pm$7.2   &  \textbf{87.5$\pm$8.8}   &  90.1$\pm$7.3   & 0.021$\pm$0.012    & 93.2$\pm$1.9     \\ \cline{1-2}\cline{4-14}
+ AD  & 411.8   &  & 94.9$\pm$6.9  & 88.3$\pm$8.2   &  91.8$\pm$6.2   & 0.029$\pm$0.014    & 92.8$\pm$1.4    &  0.66    & 93.8$\pm$7.9  & 85.7$\pm$8.9   &  89.9$\pm$7.1   & 0.021$\pm$0.012    & 93.3$\pm$1.4     \\ \cline{1-2}\cline{4-14}
+ DS \cite{zhu2017deeply} &  412.4  &   & 94.8$\pm$7.1   & 87.6$\pm$8.8   &  91.7$\pm$6.2   & 0.027$\pm$0.013    & 93.1$\pm$1.7   & 0.65   &      93.9$\pm$7.6   & 85.9$\pm$9.4   &  90.3$\pm$6.9   & 0.021$\pm$0.011    & 93.5$\pm$1.6    \\ \cline{1-2}\cline{4-14}
+ DS \cite{zhu2017deeply} + FR  &  423.7 &   & 96.0$\pm$5.3   & 89.9$\pm$7.3   &  92.7$\pm$5.7   & 0.031$\pm$0.013    & 92.9$\pm$1.6   &   0.72    &   94.2$\pm$6.8   &    86.8$\pm$8.7   &    90.4$\pm$7.2   & 0.021$\pm$0.011    &  \textbf{93.5$\pm$1.5}    \\ \cline{1-2}\cline{4-14}
Our proposed & 423.1 &   & \textbf{96.2$\pm$5.8} & 90.7$\pm$6.9  & \textbf{93.6$\pm$5.0}   &  0.035$\pm$0.014   &  92.5$\pm$2.0   &   0.77     &  \textbf{94.3$\pm$6.6}      &    86.7$\pm$8.5   & \textbf{90.6$\pm$6.7}    &  \textbf{0.021$\pm$0.011}   &  93.5$\pm$1.6    \\ \hline
\end{tabular}
}
\end{table*}

\subsubsection{Attention Distillation} In Table \ref{table:abaw}, under the same threshold $th=0.5$, AD respectively improved baseline in BD, TD, and TPR by 3.3\%, 7.0\%, and 4.6\%. It exceeds DS \cite{zhu2017deeply} in BD, TD, and TPR no matter whether FR is introduced or not. Under the same FPR, DS \cite{zhu2017deeply} alone performed slightly better than AD. When FR was combined, AD gained a slim advantage over DS \cite{zhu2017deeply} in BD and TPR.

In Table \ref{table:abav}, consistent improvements with 1.7\% of mean ACC, 1.1\% of median ACC, 1.6\% of TPR, 0.003\% of FPR, 1.2\% of DSC, 1.2\% of BD, and 0.9\% of TD were observed in AD with regard to baseline + AP (proposed). DS \cite{zhu2017deeply} also boosted performance of baseline + AP (proposed) with 0.9\% of mean ACC, 0.4\% of median ACC, 0.9\% of TPR, 0.6\% of DSC, 0.3\% of BD, and 0.2\% of TD. Moreover, AD surpassed DS \cite{zhu2017deeply} in all metrics regardless of the presence of FR.

\begin{table*}[htbp]
\centering
\caption{{Results of ablation study on pulmonary artery-vein segmentation. cSE = channel-Squeeze-Excitation, PE = Project-Excitation, FR = Feature Recalibration, AD = Attention Distillation, DS = Deep Supervision, AP = Anatomy Prior}}
\label{table:abav}
\scalebox{0.76}{
\begin{tabular}{l|l|l|l|l|l|l|l|l}
\hline
Method    &  Params ($\times10^4$)       & ACC-mean {[}95\%-CI{]} (\%) & ACC-median {[}95\%-CI{]} (\%)  & TPR  (\%) & FPR (\%) & DSC (\%) & BD (\%) & TD (\%) \\ \hline
Baseline & 1646.9 & 87.0 [84.1,89.8]  &  87.7 [84.4,92.4]  & 86.7$\pm$3.8   &  \textbf{0.130$\pm$0.037}   &  82.1$\pm$2.9  & 81.3$\pm$6.1 &  87.6$\pm$4.5   \\ \hline
+ AP (proposed) & \multirow{5}{*}{1647.1}  & 88.1 [84.5,91.7]  & 89.8 [84.9,93.1]   &  88.1$\pm$4.8  &  0.147$\pm$0.026   &  81.4$\pm$2.8  & 83.6$\pm$6.1  & 89.4$\pm$4.6   \\ \cline{1-1}\cline{3-9}
+ AP (3-D U-Net \cite{cciccek20163d}) &    & 87.9 [84.1,91.7]  &  90.1 [84.5,92.7]  & 87.9$\pm$5.1  &  0.151$\pm$0.026  &  81.0$\pm$3.0  &  83.6$\pm$5.9  &  89.2$\pm$4.6  \\ \cline{1-1}\cline{3-9}
+ AP (V-Net \cite{milletari2016v}) &     & 88.0 [84.3,91.6]  &  90.1 [84.6,92.7]  &  88.0$\pm$4.9  & 0.149$\pm$0.026  &  81.1$\pm$2.8  &  83.7$\pm$5.9  &  89.3$\pm$4.5   \\ \cline{1-1}\cline{3-9}
+ AP (VoxResNet \cite{chen2018voxresnet}) &     &  88.0 [84.4,91.6]  &  90.0 [84.6,92.8]  &  88.0$\pm$4.9 &  0.152$\pm$0.027 &  81.0$\pm$2.7  &  83.7$\pm$5.8  & 89.3$\pm$4.5  \\ \cline{1-1}\cline{3-9}
{+ AP (AG U-Net \cite{schlemper2019attention})}     &     & {88.0 [84.3,91.7]}      &     {90.0 [84.7,92.7]}        &    {88.0$\pm$4.9}  &  {0.148$\pm$0.027}      &    {81.2$\pm$2.8}      &  {83.7$\pm$5.9}      &  {89.3$\pm$4.6}      \\ \hline
+ AP + cSE \cite{zhu2019anatomynet} & 1690.8 & 88.8 [85.6,92.0]  &  89.1 [85.4,94.2]  &  88.8$\pm$4.3  &  0.139$\pm$0.039   &  82.4$\pm$3.2    & 83.3$\pm$6.6  & 89.4$\pm$4.8  \\ \hline
+ AP + PE \cite{rickmann2019project} & 1690.8  &  89.5 [86.4,92.7]  &  90.1 [86.4,94.1]  &  89.5$\pm$4.2   &  0.155$\pm$0.049 & 81.7$\pm$3.5   &  84.3$\pm$6.5  &  90.1$\pm$4.7  \\ \hline
+ AP + FR & 1691.0 & 89.4 [86.5,92.3]  & 90.2 [86.1,94.3]   &  89.4$\pm$3.8  &  0.150$\pm$0.044   &  82.0$\pm$3.5   &  83.8$\pm$5.9  &  89.9$\pm$4.2   \\ \hline
+ AP + AD &  1647.1  & 89.8 [86.9,92.7] &  90.9 [86.6,94.4]  & 89.7$\pm$3.9   &  0.144$\pm$0.043   &   \textbf{82.6$\pm$3.9}    & 84.8$\pm$5.4  &  90.3$\pm$4.2  \\ \hline
+ AP + DS \cite{zhu2017deeply} & 1647.2  & 89.0 [86.0,92.1]  &  90.2 [87.0,93.9]  & 89.0$\pm$4.1   & 0.146$\pm$0.046    &   82.0$\pm$3.8   &  83.9$\pm$5.9  &  89.6$\pm$4.4 \\ \hline
+ AP + FR + DS \cite{zhu2017deeply}   & 1691.2 & 89.7 [86.9,92.5]  & 90.5 [86.9,94.2]   & 89.7$\pm$3.7   &  0.153$\pm$0.044   &  82.0$\pm$3.2    &  84.0$\pm$5.7  &  90.1$\pm$4.2   \\ \hline
Our proposed & 1691.0 &  \textbf{90.3 [87.7,92.9]} &  \textbf{90.9 [87.4,94.6]}  &   \textbf{90.3$\pm$3.5}  &   0.151$\pm$0.043  &   82.4$\pm$3.0   &  \textbf{85.4$\pm$5.3}    &    \textbf{90.9$\pm$3.8}    \\ \hline
\end{tabular}
}
\end{table*}

\subsubsection{Anatomy Prior} Table \ref{table:abav} shows that AP (proposed) improved baseline by 1.1\% of mean ACC, 2.1\% of median ACC, 1.4\% of TPR, 2.3\% of BD, and 1.8\% of TD. AP methods that were calculated using other airway results also performed better than baseline in mean and median ACC, TPR, BD, and TD. {Among the baseline + AP methods that were computed from different airway segmentation sources, although the performance variation is small, the highest increments of mean ACC, TPR, and TD were achieved by AP (proposed) whereas AP (3-D U-Net \cite{cciccek20163d})) yielded the worst results in mean ACC, TPR, DSC, BD, and TD.}

\subsection{Qualitative Results}

\begin{figure*}[htbp]
\centerline{\includegraphics[width=\textwidth]{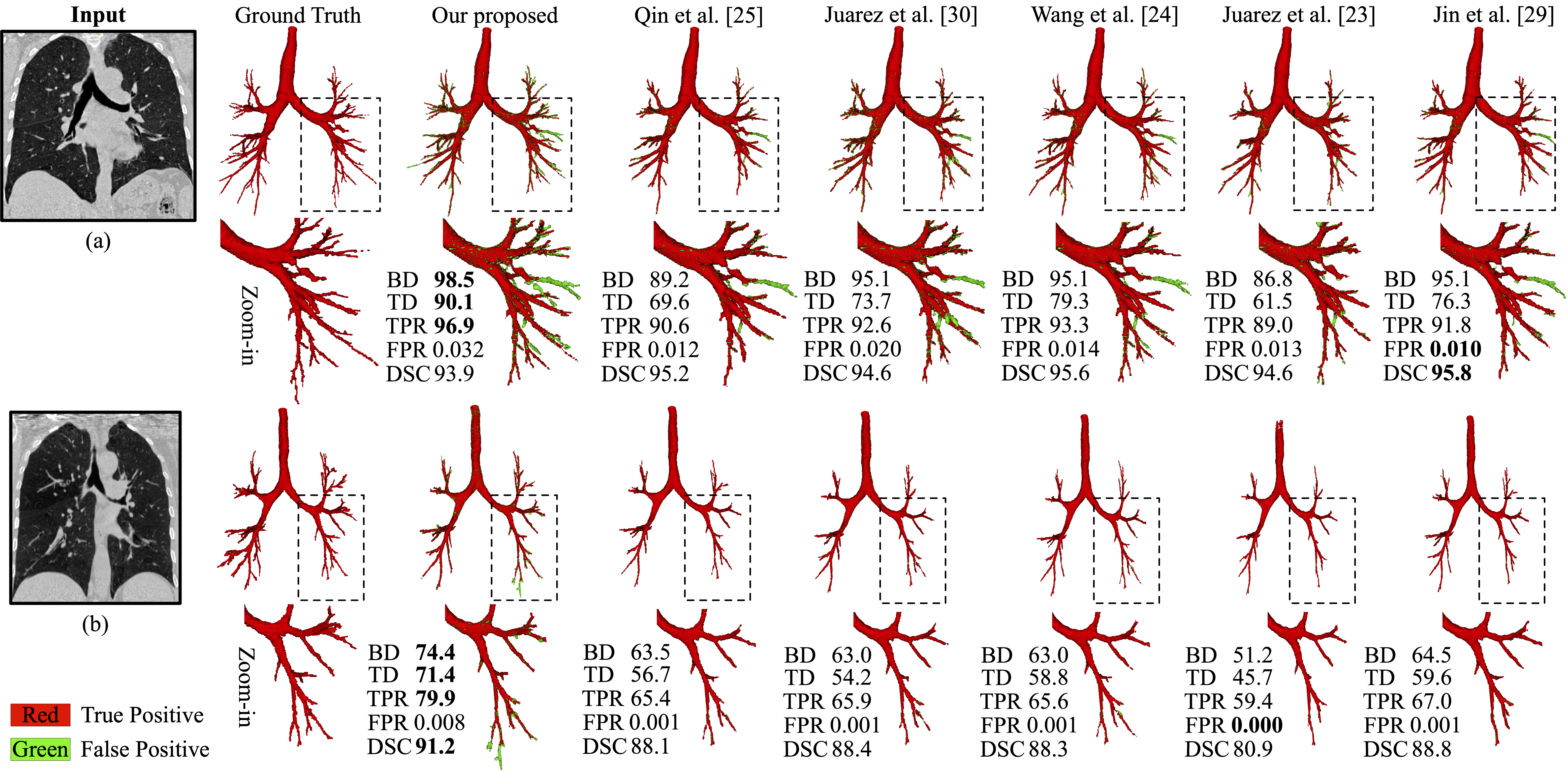}}
\caption{Rendering of pulmonary airway segmentation results on (a) easy and (b) hard testing cases. Best viewed magnified.}
\label{fig:visualaw}
\end{figure*}

\begin{figure*}[htbp]
\centerline{\includegraphics[width=\textwidth]{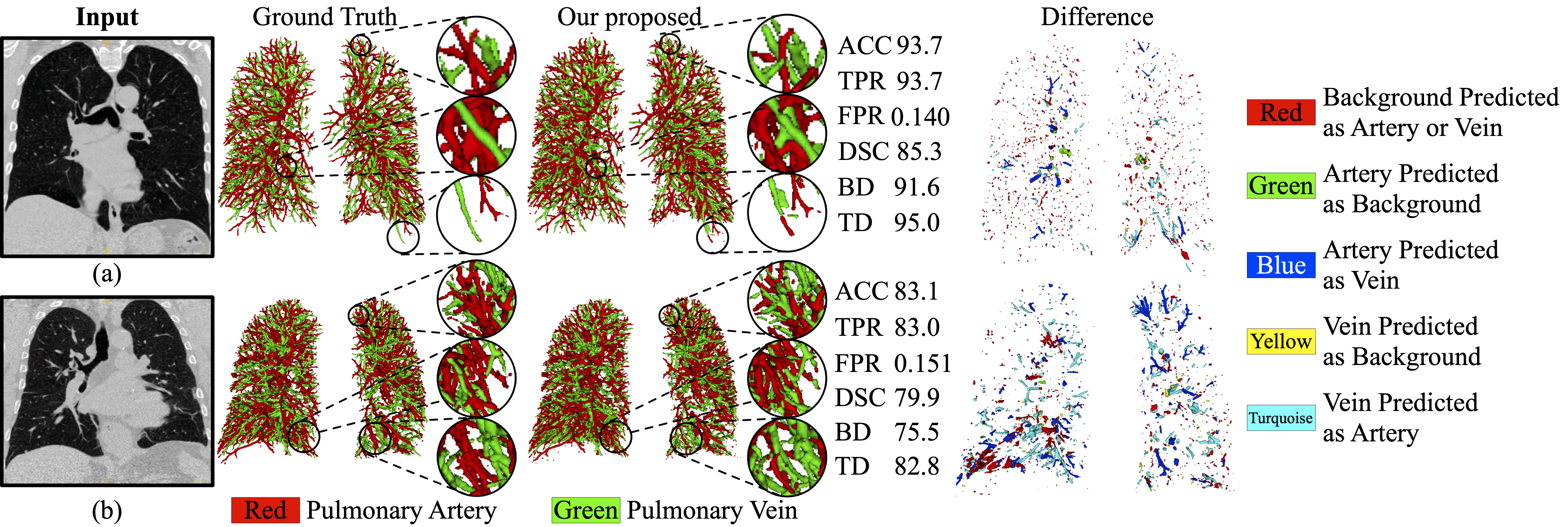}}
\caption{Rendering of pulmonary artery-vein segmentation results on (a) easy and (b) hard testing cases. Left: Wrongly segmented arteries and veins are zoomed in for better visual inspection. Right: Difference between prediction and label is categorized into 5 types.}
\label{fig:visualav}
\end{figure*}

Results of pulmonary airway and artery-vein segmentation are 3-D rendered in Fig. \ref{fig:visualaw}, illustrating the robustness of our airway segmentation method on both easy and hard cases. In line with Table \ref{table:compaw}, all methods performed well on extracting thick bronchi. Compared with state-of-the-art methods, more visible tiny branches were reconstructed by the proposed method with high overall segmentation performance maintained. Some false positives were actually true airway branches. Fig. \ref{fig:visualav} reveals that the proposed artery-vein segmentation method successfully extracted multiple arteries and veins. After close inspection of wrong predictions, we noticed that our method may fail to correctly classify some isolated vessel segments. Spatial inconsistency was also observed at terminal ends of arteries and veins (e.g., top and bottom area).

\section{Discussion}
\label{sec:discussion}

From results in Table \ref{table:compaw}, it is conclusive that our method outperformed the others in airway segmentation, especially distal thin branches. This can be ascribed to the recalibrated features and reinforced attention on hard, tiny, peripheral branches. Although CNNs possess strong fitting ability, it is necessary to suppress redundant, irrelevant features and strengthen task-related ones. Under the same threshold, two reasons are responsible for our relatively inferior FPR and DSC: 1) Our model successfully detected some true thin airways that were too indistinct to be annotated properly by experts. After careful examination of segmented airways and retrospective evaluation of labels, some branches were unintentionally neglected due to annotation difficulty. When calculating the evaluation metrics, these actually existing branches were counted as false positives and caused higher FPR with lower DSC. 2) A little leakage was produced at bifurcations when the contrast between airway lumen and wall was fairly low. In this situation, the proposed method was inclined to predict voxels as airway while other methods were relatively conservative. Some leakage regions do resemble airway in appearance, where tubular parenchyma with high-intensity circular boundary and low-intensity hollow was observed. Under the same FPR, the superior BD, TPR, and DSC of the proposed method demonstrated its sensitivity and robustness on extracting small airways. Since our threshold $th$ was increased to 0.77, airway predictions in low confidence were excluded and false positives were suppressed. The overall performance indicator DSC was consequently improved. By considering results both under the same threshold and FPR, we believe the superiority of the proposed method is well revealed.

An additional evaluation on the EXACT'09 testing set verified that under different FPR levels, our method did extract much more branches than previous methods. Besides, there exists a gap between results on our testing set and results on EXACT'09 testing set. The reasons behind are 3-fold: 1) difference in quality between our labels and EXACT'09 labels (e.g., inter-observer variation in labeling the 5-th and 6-th order bronchi); 2) difference in implementation of metrics calculation (e.g., centerline extraction); 3) difference in data distribution between the training set and EXACT'09 testing set (e.g., multi-center CT scans, dissimilar lung diseases).

Table \ref{table:compav} shows that state-of-the-art CNNs may not be effective to segment arteries and veins if no ``customization" was involved for this task. It is undeniable that two state-of-the-art methods \cite{charbonnier2015automatic, nardelli2018pulmonary} performed well in this task. In \cite{charbonnier2015automatic}, graph representation was computed and the label of artery or vein was assigned to the entire linked sub-trees. It eliminated the possibility of label inconsistency for each branch and therefore performed better than ours in terms of ACC. The two-stage CNNs-based classification method \cite{nardelli2018pulmonary} achieved the highest ACC. It highly relied on graph-cuts post-processing to refine the predictions of CNNs, where over 10\% performance gains were brought by graph-cuts. In contrast, our method is end-to-end and segmentation was directly fulfilled by CNNs. No vessel segmentation beforehand or post-processing afterwards was designed in the pipeline, which avoids error accumulation. In addition, the proposed CNNs-based method is effective for both pulmonary airway and artery-vein segmentation, which is the first of its kind in literature.

{Furthermore, post-processing experiments were performed to assess the performance gains brought by the graph-cuts \cite{boykov2001fast} method. Due to the fundamental differences between the proposed method and Ref. \cite{nardelli2018pulmonary}, the same graph-cuts technique in \cite{nardelli2018pulmonary} is not directly applicable. Modifications were made as described in Table \ref{table:compav}. The Graph-cuts (a) and (b) differ in vessel graph construction: (a) combining both the predicted arteries and veins from our CNNs as vessels; (b) combining both the ground-truth arteries and veins from labels as vessels. We observe that both Graph-cuts (a) and (b) improved artery-vein separation. Given segmented vessels, the connectivity of each voxel node with respect to its neighbor nodes is explicitly modeled in each graph. After graph construction, adjacent voxel nodes that share similar intensity tend to be in the same category. Consequently, spatial inconsistency of label assignment could be reduced for connected vessels. The comparison between Graph-cuts (a) and (b) reveals that the performance of artery-vein classification is positively associated with that of vessel segmentation. Results of Graph-cuts (b) demonstrate our superior artery-vein classification competence given ground-truth vessel branches. It is noted that the capability of graph-cuts may not be fully exploited under the current segmentation framework. Elaborate design on the incorporation of graph-cuts is needed but beyond the scope of the present study.}


To study the effectiveness of the proposed FR, we conducted extensive ablation experiments. As shown in Table \ref{table:abaw}, cSE \cite{zhu2019anatomynet}, PE \cite{rickmann2019project}, and FR all contributed to model's capacity of peripheral bronchiole extraction. Interestingly, these modules more or less worsen FPR and DSC under the same threshold. We believe the recalibration mechanism by element-wise weighting prefers airway voxels and the model tends to identify as many branches as possible, causing an increase of FPR and a decrease of DSC. Under the same FPR, the advantage of FR over baseline and other recalibration modules is clearly revealed in BD, TD, and TPR, substantiating the importance of reasonably integrating spatial knowledge for channel-wise recalibration. Table \ref{table:abav} reveals that all recalibration modules improved sensitivity of baseline (AP) to arteries and veins. The similar performance of PE \cite{rickmann2019project} and FR might be explained by the spatial distribution of artery-vein targets. Arteries and veins spread all over the lung and their difference in position may not be highly informative for artery-vein separation.

\begin{figure}[htbp]
\centerline{\includegraphics[width=\columnwidth]{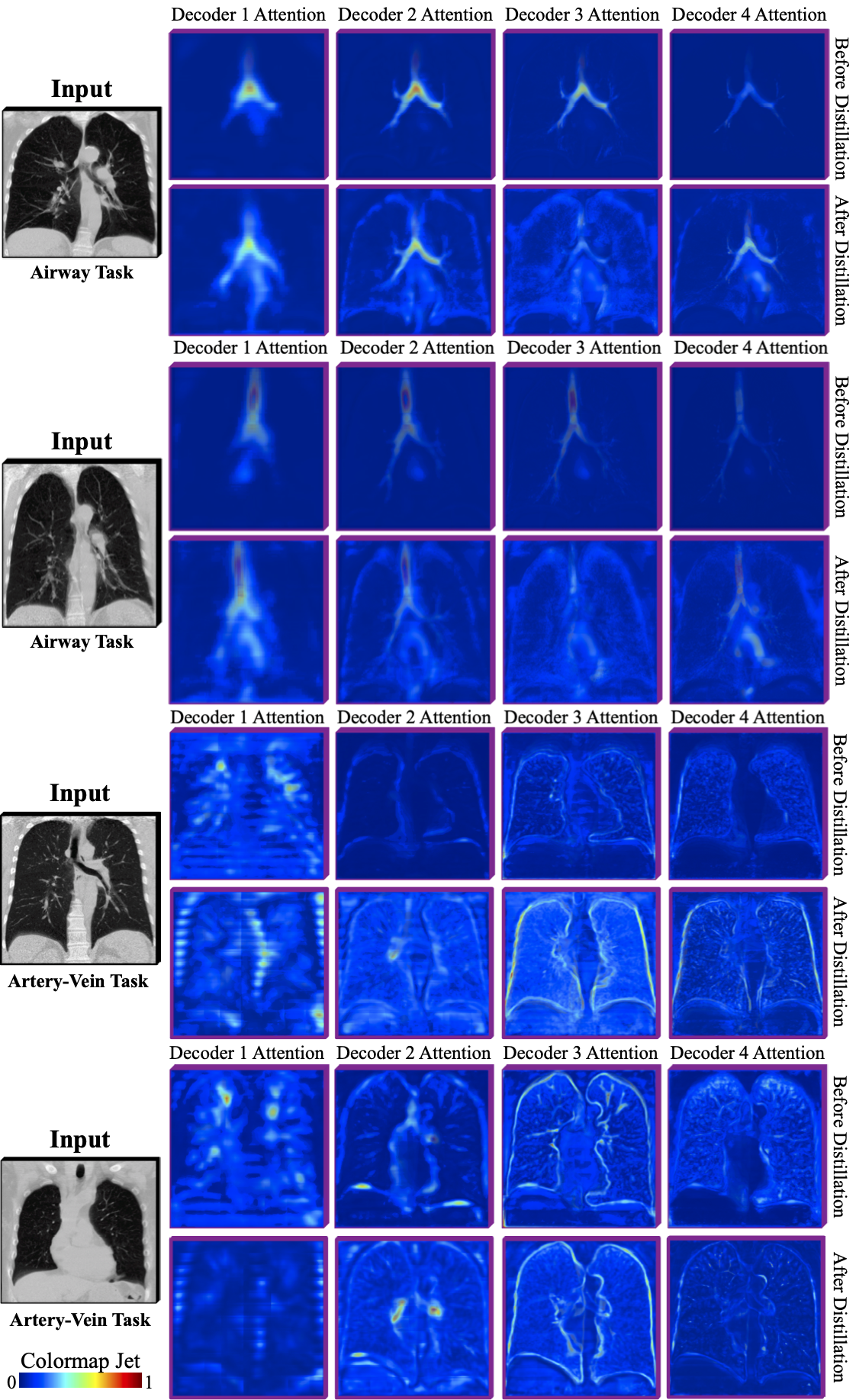}}
\caption{{Pseudo-color rendering of attention maps (decoder 1--4) before and after distillation process. These maps are min-max scaled and rendered with Jet colormap. Best viewed magnified.}}
\label{fig:attdsti}
\end{figure}

For completeness, we also conducted experiments to investigate the efficacy of the proposed AD. In Table \ref{table:abaw}, both AD and DS \cite{zhu2017deeply} improved baseline, confirming our hypothesis that it is difficult for CNNs to learn effective representation of small, thin targets only with supervision from the last output layer. In Table \ref{table:abav}, AD outperformed baseline (AP) in all metrics. Such improved sensitivity was attributed to the mechanism that features of shallower layers learned to focus on fine-grained details in features of deeper layers. To intuitively assess AD, attention maps from decoder 1--4 are visualized in Fig. \ref{fig:attdsti}. After distillation, activated regions become more distinct and the target tubules are enhanced. The improved attention on airway, vessel, and lung border explains that our model comprehended more context and therefore achieved higher sensitivity to intricate tubules. Another interesting finding is that although the last attention map is not refined in distillation, it still gets polished up because better representation learned at previous layers in turn affects late-layer features. Moreover, it is noted that AD surpassed DS \cite{zhu2017deeply} if their performance in both airway and artery-vein segmentation was comprehensively considered. It may not be optimum for earlier layers to be directly supervised by scattered and sparse targets. As shown in Fig. \ref{fig:attdsti}, not only segmentation targets but also lung contours are enhanced. DS \cite{zhu2017deeply} may hamper shallow layers from learning rich context for later comprehension.

\begin{figure}[htbp]
\centerline{\includegraphics[width=\columnwidth]{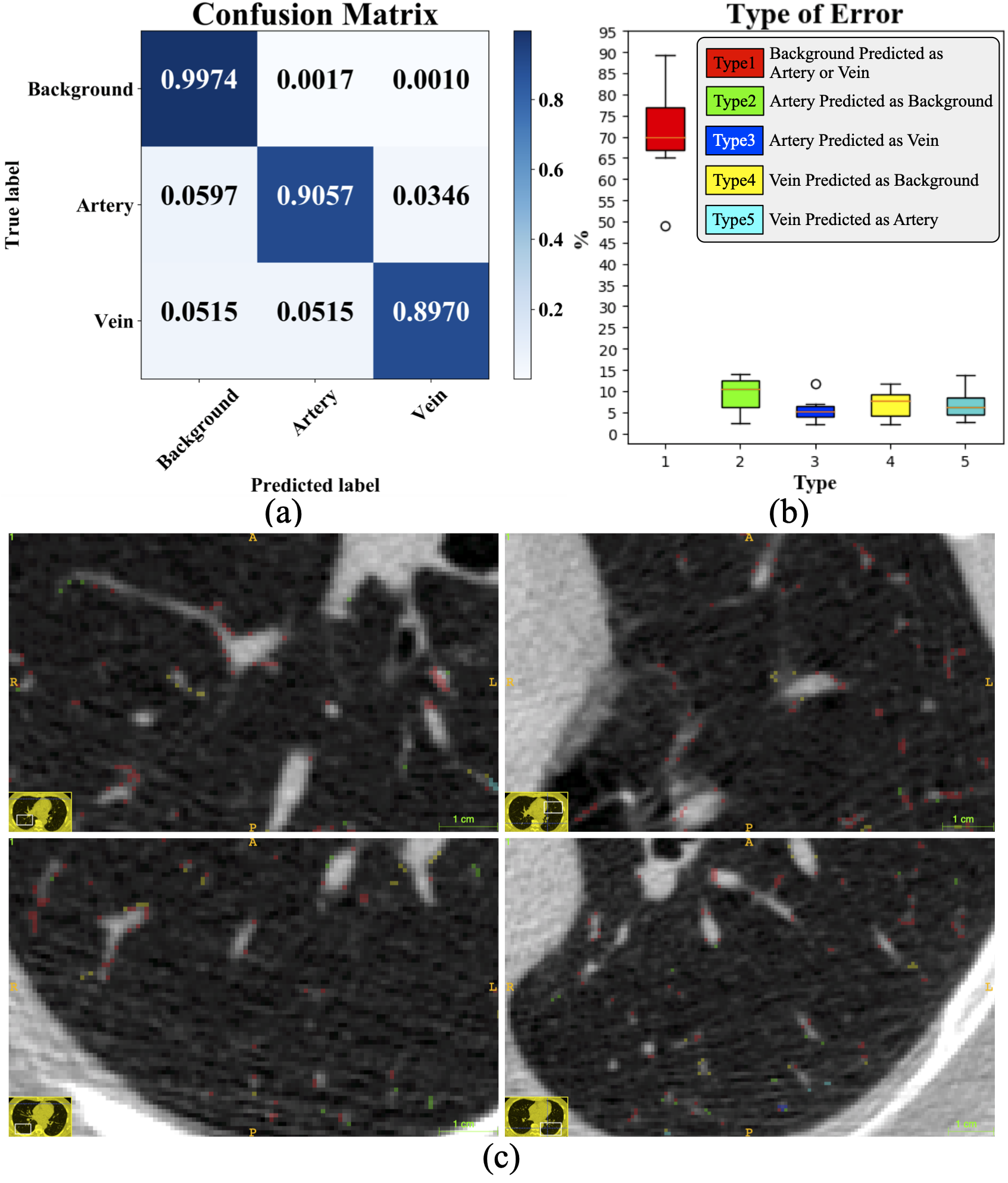}}
\caption{Analysis of multi-class artery-vein segmentation results. (a) Normalized confusion matrix. (b) Percentage of different types of errors. (c) Typical examples of wrong predictions. Best viewed magnified.}
\label{fig:avana}
\end{figure}

Ablation study on AP in Table \ref{table:abav} suggested that: 1) The lung context and distance transform maps do contribute to recognizing arteries and veins. 2) The accuracy of segmented airways was positively associated with the artery-vein segmentation performance. The more complete and precise the airway is, the more informative the calculated anatomy prior is. 3) The proposed model performed robustly to AP using different airway segmentation methods. No drastic decline was observed due to poor airway prediction results. Note that the loss of DSC by introducing AP was later offset by FR and AD.

From visual analysis, we find that some true airways were neglected unintentionally in labels (see Fig. \ref{fig:visualaw}). Such mistake was due to the weak intensity contrast and limitation of annotating 3-D objects in 2-D planes. In Fig. \ref{fig:visualav}, the reason why wrong classification was observed on isolated vessel segments and terminal ends might be explained as follows. In these regions, CNNs may not capture enough context knowledge for proper inference and no label propagation was enforced to make consistent decisions on the same vessel segment.

We further provide analysis for the artery-vein task. The confusion matrix in Fig. \ref{fig:avana}(a) shows that the proposed method performed worst on recognizing veins. Lacking anatomic relationship such as the proximity of arteries to airways, veins are a bit more difficult to segment. By dividing errors into 5 types (see Fig. \ref{fig:avana}(b)), we find that the Type 1 error of predicting background as artery or vein makes up a large proportion of all errors. In accord with Fig. \ref{fig:visualav} and Fig. \ref{fig:avana}(c), Type 1 error voxels distribute all over the lung. Besides, most errors of Type 1, 2 and 4 appear at vessel boundaries which are ambiguous for accurate and unified definition. In that case, the proposed method behaved a bit aggressively due to the reinforced learning of tubular targets.

Although the proposed method solved both pulmonary airway and artery-vein segmentation with higher sensitivity to peripheral tubular branches, there exist some limitations. First, for the current artery-vein segmentation task, only vessels inside lungs are considered as targets. The main pulmonary artery and vein vessels, including the trunk, are too difficult for human observers to delineate their boundaries in non-contrast CT. To have a broader application, CT pulmonary angiogram would be investigated in the future for segmentation of vessels in the lung hilum. Second, refinement of airway annotations might be carefully conducted to incorporate fuzzy and unclear airway branches. One efficient way of correction would be to first apply the proposed method on CT and then review both the predicted and manually labeled branches. Third, the proposed method did not enforce label compatibility in artery-vein segmentation. Spatial inconsistency was therefore observed in distal vessels. Future work includes the adoption of strategies like label propagation and majority voting. It could explicitly remove conflicting predictions that mainly caused Type 3 and 5 errors. Finally, for both airway and artery-vein tasks, more diverse clinical data might be collected to improve and examine the generalizability of our method.

\section{Conclusion}
\label{sec:conclusion}

This paper presented a tubule-sensitive method for both pulmonary airway and artery-vein segmentation. It utilizes CNNs and requires no post-processing. With the proposed spatial-aware feature recalibration module and the gradually reinforced attention distillation module, feature learning of our CNNs becomes more effective and relevant to target tubule perception. The incorporated anatomy prior is also beneficial for artery-vein separation. Extensive experiments showed that our method detected much more bronchioles, arterioles, and venules while maintaining competitive overall segmentation performance, which corroborates its superior sensitivity over state-of-the-art methods and the validity of its constituents.



\section*{Acknowledgment}
The authors are grateful to anonymous reviewers for their helpful comments. The authors would like to express gratitude to Prof. Jens Petersen and Prof. Marleen de Bruijne (DIKU) for their help in evaluating EXACT'09 testing results. The authors also thank CC-IN2P3 for their computing resources.




\bibliographystyle{IEEEtran}

\newpage


\begin{title}
\centering{\Large{---Supplementary Material---}}\\
\centering{\Large{Learning Tubule-Sensitive CNNs for}}\\
\centering{\Large{Pulmonary Airway and Artery-Vein}}\\
\centering{\Large{Segmentation in CT}}\\
\end{title}

\section{Graph-based post-processing}
\label{sec:gpp}

This section presents a detailed comparison of artery-vein segmentation before and after graph-based post-processing. Since the graph-cuts in \cite{nardelli2018pulmonary} cannot be applied to the proposed method, we consider another popular graph-based post-processing: fully connected conditional random fields (dense CRFs) \cite{krahenbuhl2011efficient}. The three-dimensional (3-D) dense CRFs \cite{kamnitsas2017efficient} was adopted to model arbitrarily large voxel-neighborhoods. In the present study, we conducted CRFs under 3 and 10 iterations. Hyper-parameters of dense CRFs are manually tuned to optimal using 6 CT scans randomly chosen from the training set. Detailed hyper-parameter settings \cite{kamnitsas2017efficient} are given as follows: $pR=12.0$, $pC=12.0$, $pZ=12.0$, $pW=1.0$, $bR=12.0$, $bC=12.0$, $bZ=12.0$, $bW=3.0$, and $bMods=10.0$. The computational time of dense CRFs is reported in Table \ref{tb:timepp}.

\begin{table}[htbp]
\centering
\caption{Computational time of the graph-based post-processing step for pulmonary artery-vein segmentation.}
\label{tb:timepp}
\scalebox{0.75}{
\begin{tabular}{l|l|l}
\hline
\multicolumn{2}{l|}{Item Name}  & Time (seconds) \\ \hline
\multirow{4}{*}{\begin{tabular}[c]{@{}l@{}}Pulmonary Artery-Vein\\ Segmentation\\ Using 55 CT Scans\end{tabular}} &  \begin{tabular}[c]{@{}l@{}}Graph-based Dense CRFs\\ Post-processing \\ 3 Iterations \\ (Per CT Volume)\end{tabular}   &  515.7$\pm$79.9 \\ \cline{2-3}
& \begin{tabular}[c]{@{}l@{}}Graph-based Dense CRFs \\ Post-processing \\ 10 Iterations \\ (Per CT Volume)\end{tabular}   &  1261.4$\pm$150.9   \\ \hline
\end{tabular}
}
\end{table}


\begin{table*}[htbp]
\centering
\caption{Results of the graph-based post-processing after the proposed artery-vein segmentation method. For each 3-D CT scan, the original intensity image and CNNs' probability outputs of background, arteries, and veins are taken as inputs to Dense CRF for post-processing. \textbf{Union 1}: The union of artery voxels before and after post-processing is kept as final artery predictions. The union of vein voxels before and after post-processing intersects with the set of non-artery voxels to obtain the final vein predictions. The remaining voxels belong to the background. \textbf{Union 2}: The union of vein voxels before and after post-processing is kept as final vein predictions. The union of artery voxels before and after post-processing intersects with the set of non-vein voxels to obtain the final artery predictions. The remaining voxels belong to the background.}
\label{tb:graph}
\scalebox{0.75}{
\begin{tabular}{l|l|l|l|l|l|l|l|l}
\hline
Method    &  Params ($\times10^4$)   &  ACC-mean {[}95\%-CI{]} (\%)   &   ACC-median {[}95\%-CI{]} (\%)   & TPR  (\%) & FPR (\%) & DSC (\%) & BD (\%) & TD (\%) \\ \hline
Our proposed & 1691.0 &  90.3 [87.7,92.9] &  90.9 [87.4,94.6]  &  90.3$\pm$3.5  &   0.151$\pm$0.043  &   82.4$\pm$3.0   &  85.4$\pm$5.3  &  90.9$\pm$3.8  \\ \hline
+ Dense CRFs (3 iterations)   & -- & 85.1 [81.7,88.4]  &  85.8 [81.2,91.5]  & 85.1$\pm$4.5  &  0.077$\pm$0.024  &  85.2$\pm$2.6  &  79.6$\pm$6.5  &  87.6$\pm$4.8  \\ \hline
+ Dense CRFs (10 iterations)   & -- & 85.0 [81.7,88.4]  &  85.7 [81.2,91.5]  & 85.1$\pm$4.5  &  0.077$\pm$0.024  &  85.2$\pm$2.6  &  79.6$\pm$6.5  &  87.6$\pm$4.8  \\ \hline
+ Dense CRFs (3 iterations)   + Union 1   & -- & 90.5 [87.9,93.1]  &  90.9 [87.6,94.6]  & 90.4$\pm$3.5  &  0.151$\pm$0.043  &  82.6$\pm$2.9  &  85.4$\pm$5.3  &  90.9$\pm$3.8  \\ \hline
+ Dense CRFs (3 iterations)   + Union 2   & -- & 90.5 [87.9,93.2]  &  91.2 [87.7,94.9]  & 90.6$\pm$3.5  &  0.151$\pm$0.043  &  82.6$\pm$2.9  &  85.7$\pm$5.3  &  91.1$\pm$3.8  \\ \hline
+ Dense CRFs (10 iterations)   + Union 1   & -- & 90.5 [87.9,93.1]  &  90.9 [87.6,94.6]  & 90.4$\pm$3.5  &  0.151$\pm$0.043  &  82.6$\pm$2.9  &  85.4$\pm$5.3  &  90.9$\pm$3.8  \\ \hline
+ Dense CRFs (10 iterations)   + Union 2   & -- & 90.5 [87.9,93.2]  &  91.2 [87.7,94.9]  & 90.6$\pm$3.5  &  0.151$\pm$0.043  &  82.6$\pm$2.9  &  85.7$\pm$5.3  &  91.1$\pm$3.8  \\ \hline
\end{tabular}
}
\end{table*}

Table \ref{tb:graph} demonstrates that after dense CRFs, the performance in mean and median ACC, TPR, BD, and TD all decreased around 3-5\%. The reasons behind are two-fold: 1) The Gibbs energy defined in CRFs imposed regularization constraints. The smoothness kernel enforced local smoothness, meaning that thin arteries and veins could be easily smoothed out by their surrounding background. Especially for peripheral branches, these scattered voxels are overwhelmed by the background majority. 2) The appearance similarity kernel enforced homogeneous appearance in CT intensity when voxels in nearby neighborhoods were identically labelled. However, there exists intraclass variance in the appearance of arteries and veins. The intensity contrast between vessel and background becomes weak and implicit around lung borders. Besides, even for human observers, it is difficult to distinguish between arteries and veins only from intensity values. The feature of CT intensity alone is unreliable for accurate inference.

The reason why FPR was reduced by half and DSC was improved by 2.8\% can also be explained by the regularization mechanism of dense CRFs. Since local smoothness and homogeneous appearance were enforced, many isolated false positives were removed and the FPR was reduced. Meanwhile, CRFs enforced connectivity within neighborhoods for thick, large vessel segments whose intensity values are similar and distinct from background. Consequently, major artery and vein segments were correctly segmented and the overall performance indicator DSC increased.

\begin{figure*}[htbp]
\centering
\includegraphics[width=\textwidth]{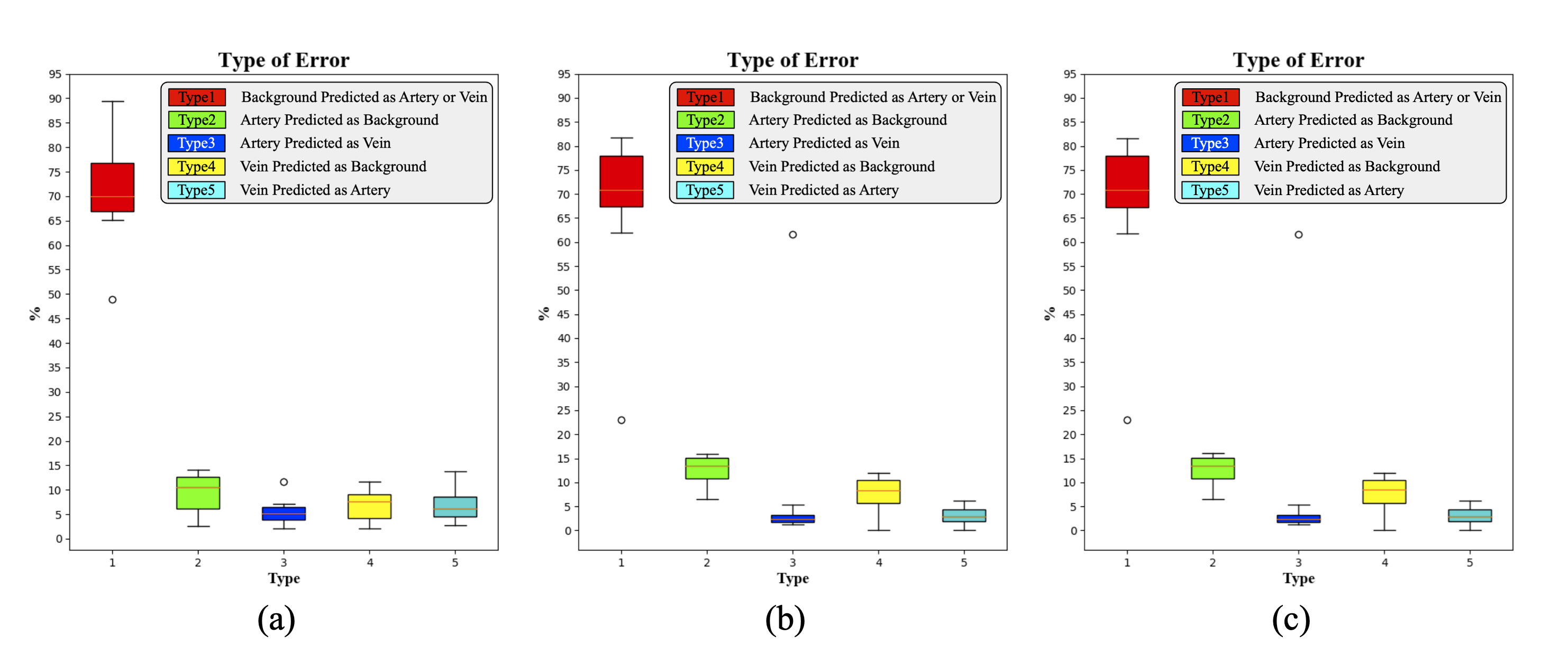}
\caption{Percentage of different types of errors by (a) the proposed method, (b) dense CRFs (3 iterations), and (c) dense CRFs (10 iterations).}
\label{fg:errorspp}
\end{figure*}

Fig. \ref{fg:errorspp} shows that compared with the proposed method, dense CRFs increased the percentage of both type 2 and type 4 errors by about 4\%. Many arteries and veins were incorrectly predicted as background, suggesting that the model's sensitivity to small, peripheral vessels was restricted. It also partly explains why performance in ACC, TPR, BD, and TD declined after dense CRFs.

The comparison of dense CRFs between 3 and 10 iterations shows that no performance gains were achieved when the number of CRFs iterations was increased.

To check if dense CRFs removed many peripheral vessels, we fused the results before and after post-processing into a union set. Arteries and veins were separately processed. To avoid overlapping between the fused arteries and veins, we used one of the following operations: 1) Union 1: intersection between the fused vein voxels and non-artery voxels; 2) Union 2: intersection between the fused artery voxels and non-vein voxels. Table \ref{tb:graph} shows that the union did ``make up" the loss of peripheral arteries and veins, recovering performance in ACC, TPR, BD, and TD. The fusing trick even improved segmentation a bit, suggesting that CRFs corrected some false predictions on thick branches (e.g., spatial inconsistency).

In summary, considering the performance and computational time, the graph-based post-processing by dense CRFs may not be suitable for the current task.

\section{Ablation study}
\label{sec:ab}

This section presents additional ablation study experiments.


\begin{table*}[htbp]
\centering
\caption{Results of ablation study on pulmonary airway segmentation. Both the results under the same binarization threshold and under the same FPR are presented for each method.}
\label{table:abawpp}
\scalebox{0.75}{
\begin{tabular}{l|l|l|l|l|l|l|l|l|l|l|l|l|l}
\hline
Method    &  Params ($\times10^4$)     & $th$  & BD (\%) & TD  (\%) & TPR (\%) & FPR (\%) & DSC (\%)  & $th$  & BD (\%) & TD  (\%) & TPR (\%) & FPR (\%) & DSC (\%)  \\ \hline
Our proposed (w/o coordinate) & 422.9   &   \multirow{14}{*}{0.5}    & 96.1$\pm$4.9 & 90.3$\pm$7.1  & 93.4$\pm$5.0   &  0.033$\pm$0.013   &  92.7$\pm$1.7 &  0.75    &    94.0$\pm$6.3    &    86.8$\pm$8.7    &    90.7$\pm$6.5    &    0.021$\pm$0.011    &    93.6$\pm$1.5  \\ \cline{1-2}\cline{4-14}
Our proposed (w/ coordinate) & 423.1  &    & 96.2$\pm$5.8 & 90.7$\pm$6.9  & 93.6$\pm$5.0   &  0.035$\pm$0.014   &  92.5$\pm$2.0    &  0.77   &  94.3$\pm$6.6      &    86.7$\pm$8.5   & 90.6$\pm$6.7    &  0.021$\pm$0.011   &  93.5$\pm$1.6    \\ \cline{1-2}\cline{4-14}
Baseline (w/o resampling) &  \multirow{2}{*}{411.8}   &    & 91.6$\pm$9.2   & 81.3$\pm$11.5   &  87.2$\pm$8.6   &  0.014$\pm$0.008    &  93.7$\pm$1.7   &   0.001   & 91.6$\pm$10.4   & 81.6$\pm$11.2   &  88.3$\pm$8.6   &  0.021$\pm$0.012    &  92.9$\pm$2.3       \\ \cline{1-1}\cline{4-14}
Baseline (w/ resampling) &     &     & 87.1$\pm$10.9   &  75.2$\pm$12.7   &  86.2$\pm$7.5   &   0.017$\pm$0.008    &   92.5$\pm$1.7      &    0.3    & 88.9$\pm$10.2    &    77.9$\pm$12.2    &    87.7$\pm$7.0    &    0.021$\pm$0.009    &    92.5$\pm$1.6    \\ \cline{1-2}\cline{4-14}
Our proposed (w/o resampling) & \multirow{10}{*}{423.1}   &    & 96.2$\pm$5.8 & 90.7$\pm$6.9  & 93.6$\pm$5.0   &  0.035$\pm$0.014   &  92.5$\pm$2.0   &   0.77     &  94.3$\pm$6.6      &    86.7$\pm$8.5   & 90.6$\pm$6.7    &  0.021$\pm$0.011   &  93.5$\pm$1.6     \\ \cline{1-1}\cline{4-14}
Our proposed (w/ resampling) &    &     &   93.6$\pm$7.3     &      87.4$\pm$9.1    & 92.3$\pm$5.5   &  0.038$\pm$0.015    &   91.4$\pm$1.9    &     0.81    &    90.3$\pm$9.2    &    80.5$\pm$11.4    &    87.7$\pm$7.1    &    0.021$\pm$0.009    &    92.3$\pm$1.6     \\ \cline{1-1}\cline{4-14}
Our proposed ($\alpha=1.0$)     &       &        &   95.3$\pm$6.3     &     88.9$\pm$7.8      &     91.9$\pm$6.1       &      0.026$\pm$0.012       &      93.3$\pm$1.6    & 0.65      &   93.6$\pm$6.7    & 86.6$\pm$8.2    &    90.4$\pm$6.7  &   0.021$\pm$0.011    &    93.6$\pm$1.5      \\ \cline{1-1}\cline{4-14}
Our proposed ($\alpha=0.5$) &   &     &   95.3$\pm$5.3     &     88.4$\pm$7.5      &       92.3$\pm$5.6       &      0.028$\pm$0.013       &      93.1$\pm$1.7   & 0.66    &   93.3$\pm$6.4    &    85.9$\pm$8.4    &    90.5$\pm$6.4    &    0.021$\pm$0.011    &    93.5$\pm$1.5      \\ \cline{1-1}\cline{4-14}
Our proposed ($\alpha=0.1$) &  &      & 96.2$\pm$5.8 & 90.7$\pm$6.9  & 93.6$\pm$5.0   &  0.035$\pm$0.014   &  92.5$\pm$2.0   &   0.77     &  94.3$\pm$6.6      &    86.7$\pm$8.5   & 90.6$\pm$6.7    &  0.021$\pm$0.011   &  93.5$\pm$1.6 \\ \cline{1-1}\cline{4-14}
Our proposed ($\alpha=0.01$) &       &     &  94.8$\pm$5.9      &        88.8$\pm$8.1        &        91.9$\pm$6.2       &       0.027$\pm$0.013       &      93.2$\pm$1.6     & 0.65 &    92.8$\pm$7.6   &  86.7$\pm$8.9    &    90.5$\pm$6.9    &  0.021$\pm$0.011    &  93.5$\pm$1.5     \\ \cline{1-1}\cline{4-14}
Our proposed ($p=10$) &      &        &    95.1$\pm$5.4      &      88.0$\pm$7.9       &      91.8$\pm$5.9        &       0.025$\pm$0.011        &       93.4$\pm$1.5     &  0.62    &    94.2$\pm$5.9    &    86.2$\pm$8.6    &    90.5$\pm$6.5    &    0.021$\pm$0.010    &    93.6$\pm$1.5      \\ \cline{1-1}\cline{4-14}
Our proposed ($p=4$) &     &      &    95.0$\pm$6.3      &      88.1$\pm$7.8       &        91.6$\pm$6.3        &       0.025$\pm$0.012        &       93.3$\pm$1.6     &    0.62    &        94.1$\pm$7.1    &    86.3$\pm$8.5    &    90.3$\pm$6.9    &    0.021$\pm$0.011    &    93.6$\pm$1.6   \\ \cline{1-1}\cline{4-14}
Our proposed ($p=2$) &     &      & 96.2$\pm$5.8 & 90.7$\pm$6.9  & 93.6$\pm$5.0   &  0.035$\pm$0.014   &  92.5$\pm$2.0  &   0.77     &  94.3$\pm$6.6      &    86.7$\pm$8.5   & 90.6$\pm$6.7    &  0.021$\pm$0.011   &  93.5$\pm$1.6     \\ \cline{1-1}\cline{4-14}
Our proposed ($p=1$) &      &       &   95.4$\pm$5.9     &     89.2$\pm$7.7      &     92.2$\pm$5.9       &      0.028$\pm$0.012      &      93.2$\pm$1.5     &   0.67   &    94.1$\pm$6.5    &    86.9$\pm$8.5    &    90.6$\pm$6.8    &    0.021$\pm$0.010    &  93.6$\pm$1.5    \\ \hline
\end{tabular}
}
\end{table*}

\begin{table*}[htbp]
\centering
\caption{Results of ablation study on pulmonary artery-vein segmentation.}
\label{table:abavpp}
\scalebox{0.75}{
\begin{tabular}{l|l|l|l|l|l|l|l|l}
\hline
Method    &   Params ($\times10^4$)   &  ACC-mean {[}95\%-CI{]} (\%)   &   ACC-median {[}95\%-CI{]} (\%)   & TPR  (\%) & FPR (\%) & DSC (\%) & BD (\%) & TD (\%) \\ \hline
Our proposed w/o auxiliary task& \multirow{2}{*}{1691.0} &    89.7 [86.2,93.2] &   91.1 [87.3,94.8]  &   89.7$\pm$4.7   &    0.158$\pm$0.048   &     81.6$\pm$4.2    &   85.2$\pm$5.9   &   90.5$\pm$4.5       \\ \cline{1-1}\cline{3-9}
Our proposed w/ auxiliary task &   &  90.3 [87.7,92.9] &  90.9 [87.4,94.6]  &  90.3$\pm$3.5  &   0.151$\pm$0.043  &   82.4$\pm$3.0   &  85.4$\pm$5.3  &  90.9$\pm$3.8  \\ \hline
Our proposed (w/o coordinate)& 1690.8 &   89.0 [85.8,92.3] &  90.2 [85.9,93.9]  &  89.0$\pm$4.3   &    0.151$\pm$0.042   &     81.7$\pm$3.9    &   83.7$\pm$6.1   &   89.5$\pm$4.7   \\ \hline
Our proposed (w/ coordinate)& \multirow{9}{*}{1691.0}  &  90.3 [87.7,92.9] &  90.9 [87.4,94.6]  &  90.3$\pm$3.5  &   0.151$\pm$0.043  &   82.4$\pm$3.0   &  85.4$\pm$5.3  &  90.9$\pm$3.8  \\ \cline{1-1}\cline{3-9}
Our proposed ($\alpha=1.0$)  &     &   89.2 [86.2,92.1] & 90.1 [86.3,94.2]  &  89.2$\pm$3.9      &    0.148$\pm$0.043      &     81.9$\pm$3.5    &      83.9$\pm$5.8   &      89.8$\pm$4.2      \\ \cline{1-1}\cline{3-9}
Our proposed ($\alpha=0.5$)  &      &    89.8 [86.9,92.7] &  90.7 [86.7,94.4]  &   89.8$\pm$3.9    &    0.158$\pm$0.043    &     81.6$\pm$3.5     &   84.4$\pm$5.8     &     90.2$\pm$4.2    \\ \cline{1-1}\cline{3-9}
Our proposed ($\alpha=0.1$)  &      &  90.3 [87.7,92.9] &  90.9 [87.4,94.6]  &  90.3$\pm$3.5  &   0.151$\pm$0.043  &   82.4$\pm$3.0   &  85.4$\pm$5.3  &  90.9$\pm$3.8  \\ \cline{1-1}\cline{3-9}
Our proposed ($\alpha=0.01$)  &      &  88.7 [85.6,91.7]   &  89.8 [85.7,93.9]  &     88.7$\pm$4.1   &    0.144$\pm$0.043    &      81.9$\pm$3.8         &      83.3$\pm$5.8    &    89.5$\pm$4.2      \\ \cline{1-1}\cline{3-9}
Our proposed ($p=10$)  &     &  88.8 [85.9,91.8] &   89.7 [85.8,93.8]  &   88.8$\pm$3.9      &       0.141$\pm$0.041      &       82.2$\pm$3.5      &      83.1$\pm$5.8   &      89.4$\pm$4.3      \\ \cline{1-1}\cline{3-9}
Our proposed ($p=4$)  &      &    89.3 [86.4,92.2] &  90.3 [85.9,94.2]  &   89.3$\pm$3.9    &    0.148$\pm$0.042    &     82.1$\pm$3.5     &   83.9$\pm$5.8     &    89.9$\pm$4.2    \\ \cline{1-1}\cline{3-9}
Our proposed ($p=2$)  &      &  90.3 [87.7,92.9] &  90.9 [87.4,94.6]  &  90.3$\pm$3.5  &   0.151$\pm$0.043  &   82.4$\pm$3.0   &  85.4$\pm$5.3  &  90.9$\pm$3.8  \\ \cline{1-1}\cline{3-9}
Our proposed ($p=1$)  &      &     88.8 [85.8,91.8]     &     89.8 [85.9,93.9]    &       88.8$\pm$4.0      &     0.140$\pm$0.042       &      82.3$\pm$3.6      &      83.3$\pm$5.9    &    89.5$\pm$4.2     \\ \hline
\end{tabular}
}
\end{table*}

\subsection{Effectiveness of Auxiliary Vessel Segmentation}
\label{sec:eavs}

To validate the effectiveness of auxiliary vessel segmentation, we removed this prediction head from CNNs' architecture and re-trained the proposed method from scratch for 70 epochs. All hyper-parameter settings remained the same.

Table \ref{table:abavpp} shows that without auxiliary vessel output, the performance declined by around 0.6\% in mean ACC, 0.6\% in TPR, 0.007\% in FPR, 0.8\% in DSC, 0.2\% in BD, and 0.4\% in TD. It demonstrates that the model's ability to differentiate between vessels and background is beneficial to separation of arteries and veins. The auxiliary vessel task provides additional gradients that may assist the optimization of CNNs. 


Although the performance gains are minor, the auxiliary vessel segmentation is effective in artery-vein segmentation.

\subsection{Effectiveness of Voxel Coordinate Map}
\label{sec:evcm}

To check effectivity of voxel coordinate map, the proposed method without coordinate information was trained from scratch and evaluated on the same dataset.

Table \ref{table:abawpp} shows that results of airway segmentation with and without coordinate information were similar. In contrast, performance of artery-vein segmentation without coordinate information degraded in almost all metrics by 0.7\%--1.7\% (Table \ref{table:abavpp}). Given limited GPU memory, since the number of parameters of the artery-vein segmentation model is much larger than that of the airway segmentation model, the input patch size for the artery-vein task ($64\times176\times176$) is smaller than that for the airway task ($80\times192\times304$). In that case, one CT sub-volume patch covers limited context and position information. The coordinates provide supplementary information about each voxel's position relative to the entire CT volume. If such coordinate map is not explicitly used, the model may not learn well the location relationship.

The voxel coordinate map did not affect much airway segmentation. Yet it did improve artery-vein segmentation.

\subsection{Negative Effects of Isometric Resampling}
\label{sec:neffect}

Isometric resampling was performed to demonstrate its negative effects on airway segmentation. We performed trilinear interpolation in CT and nearest neighbor interpolation in annotations. The resampled data share the same isotropic resolution and slice thickness of 0.625 mm. We re-trained the proposed CNNs on resampled data from scratch.

Results in Table \ref{table:abawpp} show that performance on resampled data degraded for both baseline and the proposed method. Two reasons are responsible: 1) The model was trained with annotations that mismatched their corresponding CT scans. Some voxels were not labelled correctly due to interpolation. 2) The evaluation metrics calculated with the resampled labels may also be inaccurate. Furthermore, extensive experiments demonstrated that without isometric resampling, CNNs can still learn effective representation of airways. Therefore, it is recommended not to perform isometric resampling.

\subsection{Hyper-parameter Tuning on $\alpha$}
\label{sec:hypera}

The hyper-parameter $\alpha$ typically ranges between 0 and 1. If $\alpha=1$, the attention distillation loss and segmentation loss are weighted equally in the total loss function. If $\alpha=0$, the distillation loss is set to 0 and the proposed method degenerates to the one without attention distillation. The higher the $\alpha$ is, the more emphasis the model will put on the attention distillation task instead of the segmentation task. Three new $\alpha$ values were tested: $\alpha=1.0$, $\alpha=0.5$, and $\alpha=0.01$. For each $\alpha$, the proposed airway and artery-vein segmentation method was trained from scratch and other hyper-parameters were kept the same with the original settings.

Table \ref{table:abawpp} shows that when $\alpha$ was increased or decreased from 0.1, BD, TD, and TPR of the proposed method declined for results both under the same threshold and the same FPR. Table \ref{table:abavpp} shows that when $\alpha$ was respectively increased or decreased from 0.1, mean and median ACC, TPR, DSC, BD, and TD all decreased but FPR remained similar.

In summary, we believe $\alpha$ should be tuned around 0.1 to achieve a balance between the segmentation loss and the attention distillation loss.


\subsection{Hyper-parameter Tuning on $p$}
\label{sec:hyperp}

The value of $p$ is typically set greater than or equal to 1. The higher the $p$ is, the more attention is addressed to highly activated regions in feature $A_m$ whose voxels' absolute values are greater than 1. Compared to $p=1$, the $p$-th power ($p>1$) magnifies such differentiation between targets and background. Three new values were tested: $p=1$, $p=4$, and $p=10$. For each $p$, the proposed airway and artery-vein segmentation method was trained from scratch and other hyper-parameters were kept the same with the original settings.

Table \ref{table:abawpp} shows that when $p$ was increased or decreased from 2, BD, TD, and TPR of the proposed method declined. If FPR was controlled to be the same, such performance drop was relatively slight. Table \ref{table:abavpp} shows that when $p$ was respectively increased or decreased from 2, mean and median ACC, TPR, DSC, BD, and TD all decreased but FPR got improved marginally.

In summary, $p$ should not be set too high or too low. The current $p=2$ worked well because: 1) The task-related regions are intensified properly over their surrounding background. 2) The difference in activation values between thick and thin branches is not that significant. Attention was focused on both large or small airways and vessels to avoid imbalance.


\section{Conclusion}
\label{sec:conclusion}

In this supplementary material, we presented a detailed comparison of graph-based post-processing. The dense CRFs did not perform well on artery-vein segmentation. Ablation study substantiated the effectiveness of auxiliary vessel segmentation and voxel coordinate map. Besides, isometric resampling is not encouraged. In hyper-parameter tuning experiments, the segmentation performance did vary when $\alpha$ and $p$ were respectively changed. But the variation is minor.

\bibliographystyle{IEEEtran}

\begin{thebibliography}{10}
\providecommand{\url}[1]{#1}
\csname url@samestyle\endcsname
\providecommand{\newblock}{\relax}
\providecommand{\bibinfo}[2]{#2}
\providecommand{\BIBentrySTDinterwordspacing}{\spaceskip=0pt\relax}
\providecommand{\BIBentryALTinterwordstretchfactor}{4}
\providecommand{\BIBentryALTinterwordspacing}{\spaceskip=\fontdimen2\font plus
\BIBentryALTinterwordstretchfactor\fontdimen3\font minus
  \fontdimen4\font\relax}
\providecommand{\BIBforeignlanguage}[2]{{%
\expandafter\ifx\csname l@#1\endcsname\relax
\typeout{** WARNING: IEEEtran.bst: No hyphenation pattern has been}%
\typeout{** loaded for the language `#1'. Using the pattern for}%
\typeout{** the default language instead.}%
\else
\language=\csname l@#1\endcsname
\fi
#2}}
\providecommand{\BIBdecl}{\relax}
\BIBdecl

\bibitem{howling1998significance}
S.~Howling, T.~Evans, and D.~Hansell, ``The significance of bronchial
  dilatation on ct in patients with adult respiratory distress syndrome,''
  \emph{Clinical Radiology}, vol.~53, no.~2, pp. 105--109, 1998.

\bibitem{shaw2002role}
R.~Shaw, R.~Djukanovic, D.~Tashkin, A.~Millar, R.~Du~Bois, and P.~Corris, ``The
  role of small airways in lung disease,'' \emph{Respiratory Medicine},
  vol.~96, no.~2, pp. 67--80, 2002.

\bibitem{fetita2004pulmonary}
C.~I. Fetita, F.~Pr{\^e}teux, C.~Beigelman-Aubry, and P.~Grenier, ``Pulmonary
  airways: {3-D} reconstruction from multislice {CT} and clinical
  investigation,'' \emph{IEEE Transactions on Medical Imaging}, vol.~23,
  no.~11, pp. 1353--1364, 2004.

\bibitem{li2019application}
Y.~Li, Y.~Dai, X.~Duan, W.~Zhang, Y.~Guo, and J.~Wang, ``Application of
  automated bronchial {3D-CT} measurement in pulmonary contusion complicated
  with acute respiratory distress syndrome,'' \emph{Journal of X-ray Science
  and Technology}, vol.~27, no.~4, pp. 641--654, 2019.

\bibitem{wu2019computed}
X.~Wu, G.~H. Kim, M.~L. Salisbury, D.~Barber, B.~J. Bartholmai, K.~K. Brown,
  C.~S. Conoscenti, J.~De~Backer, K.~R. Flaherty, J.~F. Gruden \emph{et~al.},
  ``Computed tomographic biomarkers in idiopathic pulmonary fibrosis. the
  future of quantitative analysis,'' \emph{American Journal of Respiratory and
  Critical Care Medicine}, vol. 199, no.~1, pp. 12--21, 2019.

\bibitem{mori2000automated}
K.~Mori, J.-I. Hasegawa, Y.~Suenaga, and J.-I. Toriwaki, ``Automated anatomical
  labeling of the bronchial branch and its application to the virtual
  bronchoscopy system,'' \emph{IEEE Transactions on Medical Imaging}, vol.~19,
  no.~2, pp. 103--114, 2000.

\bibitem{natori2005virtual}
H.~Natori, H.~Takabatake, M.~Mori, H.~Koba, K.~Mori, T.~Kitasaka, and
  Y.~Suenaga, ``Virtual navigation of central and peripheral airways,''
  \emph{Chest}, vol. 128, no.~4, p. 327S, 2005.

\bibitem{shen2015robust}
M.~Shen, S.~Giannarou, and G.-Z. Yang, ``Robust camera localisation with depth
  reconstruction for bronchoscopic navigation,'' \emph{International Journal of
  Computer Assisted Radiology and Surgery}, vol.~10, no.~6, pp. 801--813, 2015.

\bibitem{shen2019context}
M.~Shen, Y.~Gu, N.~Liu, and G.-Z. Yang, ``Context-aware depth and pose
  estimation for bronchoscopic navigation,'' \emph{IEEE Robotics and Automation
  Letters}, vol.~4, no.~2, pp. 732--739, 2019.

\bibitem{melot2011pulmonary}
C.~Melot and R.~Naeije, ``Pulmonary vascular diseases,'' \emph{Comprehensive
  Physiology}, vol.~1, no.~2, pp. 593--619, 2011.

\bibitem{charbonnier2015automatic}
J.-P. Charbonnier, M.~Brink, F.~Ciompi, E.~T. Scholten, C.~M. Schaefer-Prokop,
  and E.~M. Van~Rikxoort, ``Automatic pulmonary artery-vein separation and
  classification in computed tomography using tree partitioning and peripheral
  vessel matching,'' \emph{IEEE Transactions on Medical Imaging}, vol.~35,
  no.~3, pp. 882--892, 2015.

\bibitem{zhou2007automatic}
C.~Zhou, H.-P. Chan, B.~Sahiner, L.~M. Hadjiiski, A.~Chughtai, S.~Patel,
  J.~Wei, J.~Ge, P.~N. Cascade, and E.~A. Kazerooni, ``Automatic multiscale
  enhancement and segmentation of pulmonary vessels in {CT} pulmonary
  angiography images for {CAD} applications,'' \emph{Medical Physics}, vol.~34,
  no.~12, pp. 4567--4577, 2007.

\bibitem{wittenberg2012acute}
R.~Wittenberg, F.~H. Berger, J.~F. Peters, M.~Weber, F.~van Hoorn, L.~F.
  Beenen, M.~M. van Doorn, J.~van Schuppen, I.~A. Zijlstra, M.~Prokop
  \emph{et~al.}, ``Acute pulmonary embolism: effect of a computer-assisted
  detection prototype on diagnosis—an observer study,'' \emph{Radiology},
  vol. 262, no.~1, pp. 305--313, 2012.

\bibitem{cartin2013pulmonary}
R.~Cartin-Ceba, K.~L. Swanson, and M.~J. Krowka, ``Pulmonary arteriovenous
  malformations,'' \emph{Chest}, vol. 144, no.~3, pp. 1033--1044, 2013.

\bibitem{estepar2013computed}
R.~S.~J. Est{\'e}par, G.~L. Kinney, J.~L. Black-Shinn, R.~P. Bowler, G.~L.
  Kindlmann, J.~C. Ross, R.~Kikinis, M.~K. Han, C.~E. Come, A.~A. Diaz
  \emph{et~al.}, ``Computed tomographic measures of pulmonary vascular
  morphology in smokers and their clinical implications,'' \emph{American
  Journal of Respiratory and Critical Care Medicine}, vol. 188, no.~2, pp.
  231--239, 2013.

\bibitem{rahaghi2016pulmonary}
F.~Rahaghi, J.~Ross, M.~Agarwal, G.~Gonz{\'a}lez, C.~Come, A.~Diaz,
  G.~Vegas-S{\'a}nchez-Ferrero, A.~Hunsaker, R.~S.~J. Est{\'e}par, A.~Waxman
  \emph{et~al.}, ``Pulmonary vascular morphology as an imaging biomarker in
  chronic thromboembolic pulmonary hypertension,'' \emph{Pulmonary
  Circulation}, vol.~6, no.~1, pp. 70--81, 2016.

\bibitem{porres2013learning}
D.~V. Porres, {\'O}.~P. Morenza, E.~Pallisa, A.~Roque, J.~Andreu, and
  M.~Mart{\'\i}nez, ``Learning from the pulmonary veins,''
  \emph{Radiographics}, vol.~33, no.~4, pp. 999--1022, 2013.

\bibitem{cochran2001trends}
S.~T. Cochran, K.~Bomyea, and J.~W. Sayre, ``Trends in adverse events after iv
  administration of contrast media,'' \emph{American Journal of Roentgenology},
  vol. 176, no.~6, pp. 1385--1388, 2001.

\bibitem{loh2010delayed}
S.~Loh, S.~Bagheri, R.~W. Katzberg, M.~A. Fung, and C.-S. Li, ``Delayed adverse
  reaction to contrast-enhanced ct: a prospective single-center study
  comparison to control group without enhancement,'' \emph{Radiology}, vol.
  255, no.~3, pp. 764--771, 2010.

\bibitem{van2009automatic}
E.~M. Van~Rikxoort, W.~Baggerman, and B.~van Ginneken, ``Automatic segmentation
  of the airway tree from thoracic {CT} scans using a multi-threshold
  approach,'' in \emph{Proceeding of the Second International Workshop on
  Pulmonary Image Analysis}, 2009, pp. 341--349.

\bibitem{lo2012extraction}
P.~Lo, B.~Van~Ginneken, J.~M. Reinhardt, T.~Yavarna, P.~A. De~Jong, B.~Irving,
  C.~Fetita, M.~Ortner, R.~Pinho, J.~Sijbers \emph{et~al.}, ``Extraction of
  airways from {CT} {(EXACT'09)},'' \emph{IEEE Transactions on Medical
  Imaging}, vol.~31, no.~11, pp. 2093--2107, 2012.

\bibitem{selvan2020graph}
R.~Selvan, T.~Kipf, M.~Welling, A.~G.-U. Juarez, J.~H. Pedersen, J.~Petersen,
  and M.~de~Bruijne, ``Graph refinement based airway extraction using
  mean-field networks and graph neural networks,'' \emph{Medical Image
  Analysis}, vol.~64, p. 101751, 2020.

\bibitem{juarez2019joint}
A.~G.-U. Juarez, R.~Selvan, Z.~Saghir, and M.~de~Bruijne, ``A joint {3D}
  {UNet}-graph neural network-based method for airway segmentation from chest
  {CTs},'' in \emph{International Workshop on Machine Learning in Medical
  Imaging}.\hskip 1em plus 0.5em minus 0.4em\relax Springer, 2019, pp.
  583--591.

\bibitem{wang2019tubular}
C.~Wang, Y.~Hayashi, M.~Oda, H.~Itoh, T.~Kitasaka, A.~F. Frangi, and K.~Mori,
  ``Tubular structure segmentation using spatial fully connected network with
  radial distance loss for {3D} medical images,'' in \emph{International
  Conference on Medical Image Computing and Computer-Assisted
  Intervention}.\hskip 1em plus 0.5em minus 0.4em\relax Springer, 2019, pp.
  348--356.

\bibitem{qin2019airwaynet}
Y.~Qin, M.~Chen, H.~Zheng, Y.~Gu, M.~Shen, J.~Yang, X.~Huang, Y.-M. Zhu, and
  G.-Z. Yang, ``{AirwayNet}: A voxel-connectivity aware approach for accurate
  airway segmentation using convolutional neural networks,'' in
  \emph{International Conference on Medical Image Computing and
  Computer-Assisted Intervention}.\hskip 1em plus 0.5em minus 0.4em\relax
  Springer, 2019, pp. 212--220.

\bibitem{yun2019improvement}
J.~Yun, J.~Park, D.~Yu, J.~Yi, M.~Lee, H.~J. Park, J.-G. Lee, J.~B. Seo, and
  N.~Kim, ``Improvement of fully automated airway segmentation on volumetric
  computed tomographic images using a 2.5 dimensional convolutional neural
  net,'' \emph{Medical Image Analysis}, vol.~51, pp. 13--20, 2019.

\bibitem{charbonnier2017improving}
J.-P. Charbonnier, E.~M. Van~Rikxoort, A.~A. Setio, C.~M. Schaefer-Prokop,
  B.~van Ginneken, and F.~Ciompi, ``Improving airway segmentation in computed
  tomography using leak detection with convolutional networks,'' \emph{Medical
  Image Analysis}, vol.~36, pp. 52--60, 2017.

\bibitem{meng2017tracking}
Q.~Meng, H.~R. Roth, T.~Kitasaka, M.~Oda, J.~Ueno, and K.~Mori, ``Tracking and
  segmentation of the airways in chest {CT} using a fully convolutional
  network,'' in \emph{International Conference on Medical Image Computing and
  Computer-Assisted Intervention}.\hskip 1em plus 0.5em minus 0.4em\relax
  Springer, 2017, pp. 198--207.

\bibitem{jin20173d}
D.~Jin, Z.~Xu, A.~P. Harrison, K.~George, and D.~J. Mollura, ``{3D}
  convolutional neural networks with graph refinement for airway segmentation
  using incomplete data labels,'' in \emph{International Workshop on Machine
  Learning in Medical Imaging}.\hskip 1em plus 0.5em minus 0.4em\relax
  Springer, 2017, pp. 141--149.

\bibitem{juarez2018automatic}
A.~G.-U. Juarez, H.~Tiddens, and M.~de~Bruijne, ``Automatic airway segmentation
  in chest {CT} using convolutional neural networks,'' in \emph{Image Analysis
  for Moving Organ, Breast, and Thoracic Images}.\hskip 1em plus 0.5em minus
  0.4em\relax Springer, 2018, pp. 238--250.

\bibitem{zhao2019bronchus}
T.~Zhao, Z.~Yin, J.~Wang, D.~Gao, Y.~Chen, and Y.~Mao, ``Bronchus segmentation
  and classification by neural networks and linear programming,'' in
  \emph{International Conference on Medical Image Computing and
  Computer-Assisted Intervention}.\hskip 1em plus 0.5em minus 0.4em\relax
  Springer, 2019, pp. 230--239.

\bibitem{buelow2005automatic}
T.~Buelow, R.~Wiemker, T.~Blaffert, C.~Lorenz, and S.~Renisch, ``Automatic
  extraction of the pulmonary artery tree from multi-slice {CT} data,'' in
  \emph{Medical Imaging 2005: Physiology, Function, and Structure from Medical
  Images}, vol. 5746.\hskip 1em plus 0.5em minus 0.4em\relax International
  Society for Optics and Photonics, 2005, pp. 730--740.

\bibitem{mekada2006pulmonary}
Y.~Mekada, S.~Nakamura, I.~Ide, H.~Murase, and H.~Otsuji, ``Pulmonary artery
  and vein classification using spatial arrangement features from {X-ray CT}
  images,'' in \emph{Proc. 7th Asia-pacific Conference on Control and
  Measurement}, 2006, pp. 232--235.

\bibitem{saha2010topomorphologic}
P.~K. Saha, Z.~Gao, S.~K. Alford, M.~Sonka, and E.~A. Hoffman,
  ``Topomorphologic separation of fused isointensity objects via multiscale
  opening: Separating arteries and veins in {3-D} pulmonary {CT},'' \emph{IEEE
  Transactions on Medical Imaging}, vol.~29, no.~3, pp. 840--851, 2010.

\bibitem{gao2012new}
Z.~Gao, R.~W. Grout, C.~Holtze, E.~A. Hoffman, and P.~Saha, ``A new paradigm of
  interactive artery/vein separation in noncontrast pulmonary {CT} imaging
  using multiscale topomorphologic opening,'' \emph{IEEE Transactions on
  Biomedical Engineering}, vol.~59, no.~11, pp. 3016--3027, 2012.

\bibitem{payer2016automated}
C.~Payer, M.~Pienn, Z.~B{\'a}lint, A.~Shekhovtsov, E.~Talakic, E.~Nagy,
  A.~Olschewski, H.~Olschewski, and M.~Urschler, ``Automated integer
  programming based separation of arteries and veins from thoracic {CT}
  images,'' \emph{Medical Image Analysis}, vol.~34, pp. 109--122, 2016.

\bibitem{nardelli2018pulmonary}
P.~Nardelli, D.~Jimenez-Carretero, D.~Bermejo-Pelaez, G.~R. Washko, F.~N.
  Rahaghi, M.~J. Ledesma-Carbayo, and R.~S.~J. Est{\'e}par, ``Pulmonary
  artery--vein classification in {CT} images using deep learning,'' \emph{IEEE
  Transactions on Medical Imaging}, vol.~37, no.~11, pp. 2428--2440, 2018.

\bibitem{rickmann2019project}
A.-M. Rickmann, A.~G. Roy, I.~Sarasua, N.~Navab, and C.~Wachinger,
  ``{‘Project \& Excite’} modules for segmentation of volumetric medical
  scans,'' in \emph{International Conference on Medical Image Computing and
  Computer-Assisted Intervention}.\hskip 1em plus 0.5em minus 0.4em\relax
  Springer, 2019, pp. 39--47.

\bibitem{zhu2019anatomynet}
W.~Zhu, Y.~Huang, L.~Zeng, X.~Chen, Y.~Liu, Z.~Qian, N.~Du, W.~Fan, and X.~Xie,
  ``{AnatomyNet}: Deep learning for fast and fully automated whole-volume
  segmentation of head and neck anatomy,'' \emph{Medical Physics}, vol.~46,
  no.~2, pp. 576--589, 2019.

\bibitem{Zagoruyko2017AT}
S.~Zagoruyko and N.~Komodakis, ``Paying more attention to attention: Improving
  the performance of convolutional neural networks via attention transfer,'' in
  \emph{Proceedings of the International Conference on Learning
  Representations}.\hskip 1em plus 0.5em minus 0.4em\relax OpenReview.net,
  2017.

\bibitem{hou2019learning}
Y.~Hou, Z.~Ma, C.~Liu, and C.~C. Loy, ``Learning lightweight lane detection
  {CNNs} by self attention distillation,'' in \emph{Proceedings of the IEEE
  International Conference on Computer Vision}.\hskip 1em plus 0.5em minus
  0.4em\relax IEEE, 2019, pp. 1013--1021.

\bibitem{boykov2001fast}
Y.~Boykov, O.~Veksler, and R.~Zabih, ``Fast approximate energy minimization via
  graph cuts,'' \emph{IEEE Transactions on Pattern Analysis and Machine
  Intelligence}, vol.~23, no.~11, pp. 1222--1239, 2001.

\bibitem{otsu1979threshold}
N.~Otsu, ``A threshold selection method from gray-level histograms,''
  \emph{IEEE Transactions on Systems, Man, and Cybernetics}, vol.~9, no.~1, pp.
  62--66, 1979.

\bibitem{van2010automatic}
E.~van Dongen and B.~van Ginneken, ``Automatic segmentation of pulmonary
  vasculature in thoracic {CT} scans with local thresholding and airway wall
  removal,'' in \emph{International Symposium on Biomedical Imaging}.\hskip 1em
  plus 0.5em minus 0.4em\relax IEEE, 2010, pp. 668--671.

\bibitem{hislop2002airway}
A.~A. Hislop, ``Airway and blood vessel interaction during lung development,''
  \emph{Journal of Anatomy}, vol. 201, no.~4, pp. 325--334, 2002.

\bibitem{miller1947lung}
W.~S. Miller, \emph{The lung}.\hskip 1em plus 0.5em minus 0.4em\relax CC
  Thomas, 1947.

\bibitem{berend1979relationship}
N.~Berend, A.~Woolcock, and G.~Marlin, ``Relationship between bronchial and
  arterial diameters in normal human lungs.'' \emph{Thorax}, vol.~34, no.~3,
  pp. 354--358, 1979.

\bibitem{kandathil2018pulmonary}
A.~Kandathil and M.~Chamarthy, ``Pulmonary vascular anatomy \& anatomical
  variants,'' \emph{Cardiovascular Diagnosis and Therapy}, vol.~8, no.~3, p.
  201, 2018.

\bibitem{cciccek20163d}
{\"O}.~{\c{C}}i{\c{c}}ek, A.~Abdulkadir, S.~S. Lienkamp, T.~Brox, and
  O.~Ronneberger, ``{3D} {U-Net}: Learning dense volumetric segmentation from
  sparse annotation,'' in \emph{International Conference on Medical Image
  Computing and Computer-Assisted Intervention}.\hskip 1em plus 0.5em minus
  0.4em\relax Springer, 2016, pp. 424--432.

\bibitem{milletari2016v}
F.~Milletari, N.~Navab, and S.-A. Ahmadi, ``{V-Net}: Fully convolutional neural
  networks for volumetric medical image segmentation,'' in \emph{International
  Conference on 3D Vision}.\hskip 1em plus 0.5em minus 0.4em\relax IEEE, 2016,
  pp. 565--571.

\bibitem{lin2017focal}
T.-Y. Lin, P.~Goyal, R.~Girshick, K.~He, and P.~Doll{\'a}r, ``Focal loss for
  dense object detection,'' in \emph{Proceedings of the IEEE International
  Conference on Computer Vision}.\hskip 1em plus 0.5em minus 0.4em\relax IEEE,
  2017, pp. 2980--2988.

\bibitem{world2001world}
W.~M. Association \emph{et~al.}, ``World medical association declaration of
  helsinki. ethical principles for medical research involving human subjects.''
  \emph{Bulletin of the World Health Organization}, vol.~79, no.~4, p. 373,
  2001.

\bibitem{armato2011lung}
S.~G. Armato~III, G.~McLennan, L.~Bidaut, M.~F. McNitt-Gray, C.~R. Meyer, A.~P.
  Reeves, B.~Zhao, D.~R. Aberle, C.~I. Henschke, E.~A. Hoffman \emph{et~al.},
  ``The lung image database consortium {(LIDC)} and image database resource
  initiative {(IDRI)}: a completed reference database of lung nodules on {CT}
  scans,'' \emph{Medical Physics}, vol.~38, no.~2, pp. 915--931, 2011.

\bibitem{py06nimg}
P.~A. Yushkevich, J.~Piven, H.~Cody~Hazlett, R.~Gimpel~Smith, S.~Ho, J.~C. Gee,
  and G.~Gerig, ``User-guided {3D} active contour segmentation of anatomical
  structures: Significantly improved efficiency and reliability,''
  \emph{NeuroImage}, vol.~31, no.~3, pp. 1116--1128, 2006.

\bibitem{lee1994building}
T.-C. Lee, R.~L. Kashyap, and C.-N. Chu, ``Building skeleton models via {3-D}
  medial surface axis thinning algorithms,'' \emph{CVGIP: Graphical Models and
  Image Processing}, vol.~56, no.~6, pp. 462--478, 1994.

\bibitem{chen2018voxresnet}
H.~Chen, Q.~Dou, L.~Yu, J.~Qin, and P.-A. Heng, ``{VoxResNet}: Deep voxelwise
  residual networks for brain segmentation from {3D} {MR} images,''
  \emph{NeuroImage}, vol. 170, pp. 446--455, 2018.

\bibitem{schlemper2019attention}
J.~Schlemper, O.~Oktay, M.~Schaap, M.~Heinrich, B.~Kainz, B.~Glocker, and
  D.~Rueckert, ``Attention gated networks: Learning to leverage salient regions
  in medical images,'' \emph{Medical Image Analysis}, vol.~53, pp. 197--207,
  2019.

\bibitem{hu2018squeeze}
J.~Hu, L.~Shen, and G.~Sun, ``Squeeze-and-excitation networks,'' in
  \emph{Proceedings of the IEEE Conference on Computer Vision and Pattern
  Recognition}, 2018, pp. 7132--7141.

\bibitem{zhu2017deeply}
Q.~Zhu, B.~Du, B.~Turkbey, P.~L. Choyke, and P.~Yan, ``Deeply-supervised {CNN}
  for prostate segmentation,'' in \emph{International Joint Conference on
  Neural Networks}.\hskip 1em plus 0.5em minus 0.4em\relax IEEE, 2017, pp.
  178--184.

\end{thebibliography}

\begin{thebibliography}{1}
\providecommand{\url}[1]{#1}
\csname url@samestyle\endcsname
\providecommand{\newblock}{\relax}
\providecommand{\bibinfo}[2]{#2}
\providecommand{\BIBentrySTDinterwordspacing}{\spaceskip=0pt\relax}
\providecommand{\BIBentryALTinterwordstretchfactor}{4}
\providecommand{\BIBentryALTinterwordspacing}{\spaceskip=\fontdimen2\font plus
\BIBentryALTinterwordstretchfactor\fontdimen3\font minus
  \fontdimen4\font\relax}
\providecommand{\BIBforeignlanguage}[2]{{%
\expandafter\ifx\csname l@#1\endcsname\relax
\typeout{** WARNING: IEEEtran.bst: No hyphenation pattern has been}%
\typeout{** loaded for the language `#1'. Using the pattern for}%
\typeout{** the default language instead.}%
\else
\language=\csname l@#1\endcsname
\fi
#2}}
\providecommand{\BIBdecl}{\relax}
\BIBdecl

\bibitem{nardelli2018pulmonary}
P.~Nardelli, D.~Jimenez-Carretero, D.~Bermejo-Pelaez, G.~R. Washko, F.~N.
  Rahaghi, M.~J. Ledesma-Carbayo, and R.~S.~J. Est{\'e}par, ``Pulmonary
  artery--vein classification in {CT} images using deep learning,'' \emph{IEEE
  Transactions on Medical Imaging}, vol.~37, no.~11, pp. 2428--2440, 2018.

\bibitem{krahenbuhl2011efficient}
P.~Kr{\"a}henb{\"u}hl and V.~Koltun, ``Efficient inference in fully connected
  {CRFs} with gaussian edge potentials,'' in \emph{Advances in Neural
  Information Processing Systems}, 2011, pp. 109--117.

\bibitem{kamnitsas2017efficient}
K.~Kamnitsas, C.~Ledig, V.~F. Newcombe, J.~P. Simpson, A.~D. Kane, D.~K. Menon,
  D.~Rueckert, and B.~Glocker, ``Efficient multi-scale {3D CNN} with fully
  connected {CRF} for accurate brain lesion segmentation,'' \emph{Medical Image
  Analysis}, vol.~36, pp. 61--78, 2017.

\end{thebibliography}



\end{document}